\documentstyle[12pt]{article}
\newcommand {\cD}{{\cal D}}

\newcommand {\cN}{{\cal N}}

\newcommand {\cW}{{\cal W}}

\def\a{\alpha}
\def\b{\beta}
\def\c{\chi}
\def\d{\delta}
\def\e{\epsilon}

\def\g{\gamma}
\def\G{\Gamma}

\def\k{\kappa}
\def\l{\lambda}
\def\m{\mu}
\def\n{\nu}

\def\q{\theta}
\def\r{\rho}

\def\t{\tau}

\def\x{\xi}

\def\L{\Lambda}

\def\P{\Pi}

\newcommand{\ad}{{\dot{\alpha}}}                           
\newcommand{\bd}{{\dot{\beta}}}                            

\newcommand{\pa}{\partial}                           
\newcommand{\hf}{\frac12}
\newcommand{\be}{\begin{equation}}
\newcommand{\ee}{\end{equation}}
\newcommand{\bea}{\begin{eqnarray}}
\newcommand{\eea}{\end{eqnarray}}
\newcommand{\non}{\nonumber}

\def \gym {{g^2_{\rm YM}}}

\def \const {{\rm const}}

\def \bi{\bibitem}
\def \ci{\cite}
\def \N {{\cal N}} \def \diag {{\rm diag}}
\def \tr {{\rm tr}}

\def\W{{\cal W}}
\def\cW{\cal W}

\renewcommand{\theequation}{\thesection.\arabic{equation}}
\def\const{{\rm const}}
\def\la{\left\langle}

\def\d{\partial}
\newcommand{\rf}[1]{(\ref{#1})}
\def \N {{\cal N }}
\def \cN {{\cal N }}
\def \del {\partial}
\def \ov {\over}
\def \la {\label}
\def\foot {\footnote}

\begin{document}
\immediate\write16{<<WARNING: LINEDRAW macros work with emTeX-dvivers
                    and other drivers supporting emTeX \special's
                    (dviscr, dvihplj, dvidot, dvips, dviwin, etc.) >>}
\newdimen\Lengthunit       \Lengthunit  = 1.5cm
\newcount\Nhalfperiods     \Nhalfperiods= 9
\newcount\magnitude        \magnitude = 1000

\catcode`\*=11
\newdimen\L*   \newdimen\d*   \newdimen\d**
\newdimen\dm*  \newdimen\dd*  \newdimen\dt*
\newdimen\a*   \newdimen\b*   \newdimen\c*
\newdimen\a**  \newdimen\b**
\newdimen\xL*  \newdimen\yL*
\newdimen\rx*  \newdimen\ry*
\newdimen\tmp* \newdimen\linwid*

\newcount\k*   \newcount\l*   \newcount\m*
\newcount\k**  \newcount\l**  \newcount\m**
\newcount\n*   \newcount\dn*  \newcount\r*
\newcount\N*   \newcount\*one \newcount\*two  \*one=1 \*two=2
\newcount\*ths \*ths=1000
\newcount\angle*  \newcount\q*  \newcount\q**
\newcount\angle** \angle**=0
\newcount\sc*     \sc*=0

\newtoks\cos*  \cos*={1}
\newtoks\sin*  \sin*={0}

\catcode`\[=13

\def\rotate(#1){\advance\angle**#1\angle*=\angle**
\q**=\angle*\ifnum\q**<0\q**=-\q**\fi
\ifnum\q**>360\q*=\angle*\divide\q*360\multiply\q*360\advance\angle*-\q*\fi
\ifnum\angle*<0\advance\angle*360\fi\q**=\angle*\divide\q**90\q**=\q**
\def\sgcos*{+}\def\sgsin*{+}\relax
\ifcase\q**\or
 \def\sgcos*{-}\def\sgsin*{+}\or
 \def\sgcos*{-}\def\sgsin*{-}\or
 \def\sgcos*{+}\def\sgsin*{-}\else\fi
\q*=\q**
\multiply\q*90\advance\angle*-\q*
\ifnum\angle*>45\sc*=1\angle*=-\angle*\advance\angle*90\else\sc*=0\fi
\def[##1,##2]{\ifnum\sc*=0\relax
\edef\cs*{\sgcos*.##1}\edef\sn*{\sgsin*.##2}\ifcase\q**\or
 \edef\cs*{\sgcos*.##2}\edef\sn*{\sgsin*.##1}\or
 \edef\cs*{\sgcos*.##1}\edef\sn*{\sgsin*.##2}\or
 \edef\cs*{\sgcos*.##2}\edef\sn*{\sgsin*.##1}\else\fi\else
\edef\cs*{\sgcos*.##2}\edef\sn*{\sgsin*.##1}\ifcase\q**\or
 \edef\cs*{\sgcos*.##1}\edef\sn*{\sgsin*.##2}\or
 \edef\cs*{\sgcos*.##2}\edef\sn*{\sgsin*.##1}\or
 \edef\cs*{\sgcos*.##1}\edef\sn*{\sgsin*.##2}\else\fi\fi
\cos*={\cs*}\sin*={\sn*}\global\edef\gcos*{\cs*}\global\edef\gsin*{\sn*}}\relax
\ifcase\angle*[9999,0]\or
[999,017]\or[999,034]\or[998,052]\or[997,069]\or[996,087]\or
[994,104]\or[992,121]\or[990,139]\or[987,156]\or[984,173]\or
[981,190]\or[978,207]\or[974,224]\or[970,241]\or[965,258]\or
[961,275]\or[956,292]\or[951,309]\or[945,325]\or[939,342]\or
[933,358]\or[927,374]\or[920,390]\or[913,406]\or[906,422]\or
[898,438]\or[891,453]\or[882,469]\or[874,484]\or[866,499]\or
[857,515]\or[848,529]\or[838,544]\or[829,559]\or[819,573]\or
[809,587]\or[798,601]\or[788,615]\or[777,629]\or[766,642]\or
[754,656]\or[743,669]\or[731,681]\or[719,694]\or[707,707]\or
\else[9999,0]\fi}

\catcode`\[=12

\def\GRAPH(hsize=#1)#2{\hbox to #1\Lengthunit{#2\hss}}

\def\Linewidth#1{\global\linwid*=#1\relax
\global\divide\linwid*10\global\multiply\linwid*\mag
\global\divide\linwid*100\special{em:linewidth \the\linwid*}}

\Linewidth{.4pt}
\def\sm*{\special{em:moveto}}
\def\sl*{\special{em:lineto}}
\let\moveto=\sm*
\let\lineto=\sl*
\newbox\spm*   \newbox\spl*
\setbox\spm*\hbox{\sm*}
\setbox\spl*\hbox{\sl*}

\def\mov#1(#2,#3)#4{\rlap{\L*=#1\Lengthunit
\xL*=#2\L* \yL*=#3\L*
\xL*=\xscale\xL* \yL*=\yscale\yL*
\rx* \the\cos*\xL* \tmp* \the\sin*\yL* \advance\rx*-\tmp*
\ry* \the\cos*\yL* \tmp* \the\sin*\xL* \advance\ry*\tmp*
\kern\rx*\raise\ry*\hbox{#4}}}

\def\rmov*(#1,#2)#3{\rlap{\xL*=#1\yL*=#2\relax
\rx* \the\cos*\xL* \tmp* \the\sin*\yL* \advance\rx*-\tmp*
\ry* \the\cos*\yL* \tmp* \the\sin*\xL* \advance\ry*\tmp*
\kern\rx*\raise\ry*\hbox{#3}}}

\def\lin#1(#2,#3){\rlap{\sm*\mov#1(#2,#3){\sl*}}}

\def\arr*(#1,#2,#3){\rmov*(#1\dd*,#1\dt*){\sm*
\rmov*(#2\dd*,#2\dt*){\rmov*(#3\dt*,-#3\dd*){\sl*}}\sm*
\rmov*(#2\dd*,#2\dt*){\rmov*(-#3\dt*,#3\dd*){\sl*}}}}

\def\arrow#1(#2,#3){\rlap{\lin#1(#2,#3)\mov#1(#2,#3){\relax
\d**=-.012\Lengthunit\dd*=#2\d**\dt*=#3\d**
\arr*(1,10,4)\arr*(3,8,4)\arr*(4.8,4.2,3)}}}

\def\arrlin#1(#2,#3){\rlap{\L*=#1\Lengthunit\L*=.5\L*
\lin#1(#2,#3)\rmov*(#2\L*,#3\L*){\arrow.1(#2,#3)}}}

\def\dasharrow#1(#2,#3){\rlap{{\Lengthunit=0.9\Lengthunit
\dashlin#1(#2,#3)\mov#1(#2,#3){\sm*}}\mov#1(#2,#3){\sl*
\d**=-.012\Lengthunit\dd*=#2\d**\dt*=#3\d**
\arr*(1,10,4)\arr*(3,8,4)\arr*(4.8,4.2,3)}}}

\def\clap#1{\hbox to 0pt{\hss #1\hss}}

\def\ind(#1,#2)#3{\rlap{\L*=.1\Lengthunit
\xL*=#1\L* \yL*=#2\L*
\rx* \the\cos*\xL* \tmp* \the\sin*\yL* \advance\rx*-\tmp*
\ry* \the\cos*\yL* \tmp* \the\sin*\xL* \advance\ry*\tmp*
\kern\rx*\raise\ry*\hbox{\lower2pt\clap{$#3$}}}}

\def\sh*(#1,#2)#3{\rlap{\dm*=\the\n*\d**
\xL*=\xscale\dm* \yL*=\yscale\dm* \xL*=#1\xL* \yL*=#2\yL*
\rx* \the\cos*\xL* \tmp* \the\sin*\yL* \advance\rx*-\tmp*
\ry* \the\cos*\yL* \tmp* \the\sin*\xL* \advance\ry*\tmp*
\kern\rx*\raise\ry*\hbox{#3}}}

\def\calcnum*#1(#2,#3){\a*=1000sp\b*=1000sp\a*=#2\a*\b*=#3\b*
\ifdim\a*<0pt\a*-\a*\fi\ifdim\b*<0pt\b*-\b*\fi
\ifdim\a*>\b*\c*=.96\a*\advance\c*.4\b*
\else\c*=.96\b*\advance\c*.4\a*\fi
\k*\a*\multiply\k*\k*\l*\b*\multiply\l*\l*
\m*\k*\advance\m*\l*\n*\c*\r*\n*\multiply\n*\n*
\dn*\m*\advance\dn*-\n*\divide\dn*2\divide\dn*\r*
\advance\r*\dn*
\c*=\the\Nhalfperiods5sp\c*=#1\c*\ifdim\c*<0pt\c*-\c*\fi
\multiply\c*\r*\N*\c*\divide\N*10000}

\def\dashlin#1(#2,#3){\rlap{\calcnum*#1(#2,#3)\relax
\d**=#1\Lengthunit\ifdim\d**<0pt\d**-\d**\fi
\divide\N*2\multiply\N*2\advance\N*\*one
\divide\d**\N*\sm*\n*\*one\sh*(#2,#3){\sl*}\loop
\advance\n*\*one\sh*(#2,#3){\sm*}\advance\n*\*one
\sh*(#2,#3){\sl*}\ifnum\n*<\N*\repeat}}

\def\dashdotlin#1(#2,#3){\rlap{\calcnum*#1(#2,#3)\relax
\d**=#1\Lengthunit\ifdim\d**<0pt\d**-\d**\fi
\divide\N*2\multiply\N*2\advance\N*1\multiply\N*2\relax
\divide\d**\N*\sm*\n*\*two\sh*(#2,#3){\sl*}\loop
\advance\n*\*one\sh*(#2,#3){\kern-1.48pt\lower.5pt\hbox{\rm.}}\relax
\advance\n*\*one\sh*(#2,#3){\sm*}\advance\n*\*two
\sh*(#2,#3){\sl*}\ifnum\n*<\N*\repeat}}

\def\shl*(#1,#2)#3{\kern#1#3\lower#2#3\hbox{\unhcopy\spl*}}

\def\trianglin#1(#2,#3){\rlap{\toks0={#2}\toks1={#3}\calcnum*#1(#2,#3)\relax
\dd*=.57\Lengthunit\dd*=#1\dd*\divide\dd*\N*
\divide\dd*\*ths \multiply\dd*\magnitude
\d**=#1\Lengthunit\ifdim\d**<0pt\d**-\d**\fi
\multiply\N*2\divide\d**\N*\sm*\n*\*one\loop
\shl**{\dd*}\dd*-\dd*\advance\n*2\relax
\ifnum\n*<\N*\repeat\n*\N*\shl**{0pt}}}

\def\wavelin#1(#2,#3){\rlap{\toks0={#2}\toks1={#3}\calcnum*#1(#2,#3)\relax
\dd*=.23\Lengthunit\dd*=#1\dd*\divide\dd*\N*
\divide\dd*\*ths \multiply\dd*\magnitude
\d**=#1\Lengthunit\ifdim\d**<0pt\d**-\d**\fi
\multiply\N*4\divide\d**\N*\sm*\n*\*one\loop
\shl**{\dd*}\dt*=1.3\dd*\advance\n*\*one
\shl**{\dt*}\advance\n*\*one
\shl**{\dd*}\advance\n*\*two
\dd*-\dd*\ifnum\n*<\N*\repeat\n*\N*\shl**{0pt}}}

\def\w*lin(#1,#2){\rlap{\toks0={#1}\toks1={#2}\d**=\Lengthunit\dd*=-.12\d**
\divide\dd*\*ths \multiply\dd*\magnitude
\N*8\divide\d**\N*\sm*\n*\*one\loop
\shl**{\dd*}\dt*=1.3\dd*\advance\n*\*one
\shl**{\dt*}\advance\n*\*one
\shl**{\dd*}\advance\n*\*one
\shl**{0pt}\dd*-\dd*\advance\n*1\ifnum\n*<\N*\repeat}}

\def\l*arc(#1,#2)[#3][#4]{\rlap{\toks0={#1}\toks1={#2}\d**=\Lengthunit
\dd*=#3.037\d**\dd*=#4\dd*\dt*=#3.049\d**\dt*=#4\dt*\ifdim\d**>10mm\relax
\d**=.25\d**\n*\*one\shl**{-\dd*}\n*\*two\shl**{-\dt*}\n*3\relax
\shl**{-\dd*}\n*4\relax\shl**{0pt}\else
\ifdim\d**>5mm\d**=.5\d**\n*\*one\shl**{-\dt*}\n*\*two
\shl**{0pt}\else\n*\*one\shl**{0pt}\fi\fi}}

\def\d*arc(#1,#2)[#3][#4]{\rlap{\toks0={#1}\toks1={#2}\d**=\Lengthunit
\dd*=#3.037\d**\dd*=#4\dd*\d**=.25\d**\sm*\n*\*one\shl**{-\dd*}\relax
\n*3\relax\sh*(#1,#2){\xL*=\xscale\dd*\yL*=\yscale\dd*
\kern#2\xL*\lower#1\yL*\hbox{\sm*}}\n*4\relax\shl**{0pt}}}

\def\shl**#1{\c*=\the\n*\d**\d*=#1\relax
\a*=\the\toks0\c*\b*=\the\toks1\d*\advance\a*-\b*
\b*=\the\toks1\c*\d*=\the\toks0\d*\advance\b*\d*
\a*=\xscale\a*\b*=\yscale\b*
\rx* \the\cos*\a* \tmp* \the\sin*\b* \advance\rx*-\tmp*
\ry* \the\cos*\b* \tmp* \the\sin*\a* \advance\ry*\tmp*
\raise\ry*\rlap{\kern\rx*\unhcopy\spl*}}

\def\wlin*#1(#2,#3)[#4]{\rlap{\toks0={#2}\toks1={#3}\relax
\c*=#1\l*\c*\c*=.01\Lengthunit\m*\c*\divide\l*\m*
\c*=\the\Nhalfperiods5sp\multiply\c*\l*\N*\c*\divide\N*\*ths
\divide\N*2\multiply\N*2\advance\N*\*one
\dd*=.002\Lengthunit\dd*=#4\dd*\multiply\dd*\l*\divide\dd*\N*
\divide\dd*\*ths \multiply\dd*\magnitude
\d**=#1\multiply\N*4\divide\d**\N*\sm*\n*\*one\loop
\shl**{\dd*}\dt*=1.3\dd*\advance\n*\*one
\shl**{\dt*}\advance\n*\*one
\shl**{\dd*}\advance\n*\*two
\dd*-\dd*\ifnum\n*<\N*\repeat\n*\N*\shl**{0pt}}}

\def\wavebox#1{\setbox0\hbox{#1}\relax
\a*=\wd0\advance\a*14pt\b*=\ht0\advance\b*\dp0\advance\b*14pt\relax
\hbox{\kern9pt\relax
\rmov*(0pt,\ht0){\rmov*(-7pt,7pt){\wlin*\a*(1,0)[+]\wlin*\b*(0,-1)[-]}}\relax
\rmov*(\wd0,-\dp0){\rmov*(7pt,-7pt){\wlin*\a*(-1,0)[+]\wlin*\b*(0,1)[-]}}\relax
\box0\kern9pt}}

\def\rectangle#1(#2,#3){\relax
\lin#1(#2,0)\lin#1(0,#3)\mov#1(0,#3){\lin#1(#2,0)}\mov#1(#2,0){\lin#1(0,#3)}}

\def\dashrectangle#1(#2,#3){\dashlin#1(#2,0)\dashlin#1(0,#3)\relax
\mov#1(0,#3){\dashlin#1(#2,0)}\mov#1(#2,0){\dashlin#1(0,#3)}}

\def\waverectangle#1(#2,#3){\L*=#1\Lengthunit\a*=#2\L*\b*=#3\L*
\ifdim\a*<0pt\a*-\a*\def\x*{-1}\else\def\x*{1}\fi
\ifdim\b*<0pt\b*-\b*\def\y*{-1}\else\def\y*{1}\fi
\wlin*\a*(\x*,0)[-]\wlin*\b*(0,\y*)[+]\relax
\mov#1(0,#3){\wlin*\a*(\x*,0)[+]}\mov#1(#2,0){\wlin*\b*(0,\y*)[-]}}

\def\calcparab*{\ifnum\n*>\m*\k*\N*\advance\k*-\n*\else\k*\n*\fi
\a*=\the\k* sp\a*=10\a*\b*\dm*\advance\b*-\a*\k*\b*
\a*=\the\*ths\b*\divide\a*\l*\multiply\a*\k*
\divide\a*\l*\k*\*ths\r*\a*\advance\k*-\r*\dt*=\the\k*\L*}

\def\arcto#1(#2,#3)[#4]{\rlap{\toks0={#2}\toks1={#3}\calcnum*#1(#2,#3)\relax
\dm*=135sp\dm*=#1\dm*\d**=#1\Lengthunit\ifdim\dm*<0pt\dm*-\dm*\fi
\multiply\dm*\r*\a*=.3\dm*\a*=#4\a*\ifdim\a*<0pt\a*-\a*\fi
\advance\dm*\a*\N*\dm*\divide\N*10000\relax
\divide\N*2\multiply\N*2\advance\N*\*one
\L*=-.25\d**\L*=#4\L*\divide\d**\N*\divide\L*\*ths
\m*\N*\divide\m*2\dm*=\the\m*5sp\l*\dm*\sm*\n*\*one\loop
\calcparab*\shl**{-\dt*}\advance\n*1\ifnum\n*<\N*\repeat}}

\def\arrarcto#1(#2,#3)[#4]{\L*=#1\Lengthunit\L*=.54\L*
\arcto#1(#2,#3)[#4]\rmov*(#2\L*,#3\L*){\d*=.457\L*\d*=#4\d*\d**-\d*
\rmov*(#3\d**,#2\d*){\arrow.02(#2,#3)}}}

\def\dasharcto#1(#2,#3)[#4]{\rlap{\toks0={#2}\toks1={#3}\relax
\calcnum*#1(#2,#3)\dm*=\the\N*5sp\a*=.3\dm*\a*=#4\a*\ifdim\a*<0pt\a*-\a*\fi
\advance\dm*\a*\N*\dm*
\divide\N*20\multiply\N*2\advance\N*1\d**=#1\Lengthunit
\L*=-.25\d**\L*=#4\L*\divide\d**\N*\divide\L*\*ths
\m*\N*\divide\m*2\dm*=\the\m*5sp\l*\dm*
\sm*\n*\*one\loop\calcparab*
\shl**{-\dt*}\advance\n*1\ifnum\n*>\N*\else\calcparab*
\sh*(#2,#3){\xL*=#3\dt* \yL*=#2\dt*
\rx* \the\cos*\xL* \tmp* \the\sin*\yL* \advance\rx*\tmp*
\ry* \the\cos*\yL* \tmp* \the\sin*\xL* \advance\ry*-\tmp*
\kern\rx*\lower\ry*\hbox{\sm*}}\fi
\advance\n*1\ifnum\n*<\N*\repeat}}

\def\*shl*#1{\c*=\the\n*\d**\advance\c*#1\a**\d*\dt*\advance\d*#1\b**
\a*=\the\toks0\c*\b*=\the\toks1\d*\advance\a*-\b*
\b*=\the\toks1\c*\d*=\the\toks0\d*\advance\b*\d*
\rx* \the\cos*\a* \tmp* \the\sin*\b* \advance\rx*-\tmp*
\ry* \the\cos*\b* \tmp* \the\sin*\a* \advance\ry*\tmp*
\raise\ry*\rlap{\kern\rx*\unhcopy\spl*}}

\def\calcnormal*#1{\b**=10000sp\a**\b**\k*\n*\advance\k*-\m*
\multiply\a**\k*\divide\a**\m*\a**=#1\a**\ifdim\a**<0pt\a**-\a**\fi
\ifdim\a**>\b**\d*=.96\a**\advance\d*.4\b**
\else\d*=.96\b**\advance\d*.4\a**\fi
\d*=.01\d*\r*\d*\divide\a**\r*\divide\b**\r*
\ifnum\k*<0\a**-\a**\fi\d*=#1\d*\ifdim\d*<0pt\b**-\b**\fi
\k*\a**\a**=\the\k*\dd*\k*\b**\b**=\the\k*\dd*}

\def\wavearcto#1(#2,#3)[#4]{\rlap{\toks0={#2}\toks1={#3}\relax
\calcnum*#1(#2,#3)\c*=\the\N*5sp\a*=.4\c*\a*=#4\a*\ifdim\a*<0pt\a*-\a*\fi
\advance\c*\a*\N*\c*\divide\N*20\multiply\N*2\advance\N*-1\multiply\N*4\relax
\d**=#1\Lengthunit\dd*=.012\d**
\divide\dd*\*ths \multiply\dd*\magnitude
\ifdim\d**<0pt\d**-\d**\fi\L*=.25\d**
\divide\d**\N*\divide\dd*\N*\L*=#4\L*\divide\L*\*ths
\m*\N*\divide\m*2\dm*=\the\m*0sp\l*\dm*
\sm*\n*\*one\loop\calcnormal*{#4}\calcparab*
\*shl*{1}\advance\n*\*one\calcparab*
\*shl*{1.3}\advance\n*\*one\calcparab*
\*shl*{1}\advance\n*2\dd*-\dd*\ifnum\n*<\N*\repeat\n*\N*\shl**{0pt}}}

\def\triangarcto#1(#2,#3)[#4]{\rlap{\toks0={#2}\toks1={#3}\relax
\calcnum*#1(#2,#3)\c*=\the\N*5sp\a*=.4\c*\a*=#4\a*\ifdim\a*<0pt\a*-\a*\fi
\advance\c*\a*\N*\c*\divide\N*20\multiply\N*2\advance\N*-1\multiply\N*2\relax
\d**=#1\Lengthunit\dd*=.012\d**
\divide\dd*\*ths \multiply\dd*\magnitude
\ifdim\d**<0pt\d**-\d**\fi\L*=.25\d**
\divide\d**\N*\divide\dd*\N*\L*=#4\L*\divide\L*\*ths
\m*\N*\divide\m*2\dm*=\the\m*0sp\l*\dm*
\sm*\n*\*one\loop\calcnormal*{#4}\calcparab*
\*shl*{1}\advance\n*2\dd*-\dd*\ifnum\n*<\N*\repeat\n*\N*\shl**{0pt}}}

\def\hr*#1{\L*=\xscale\Lengthunit\ifnum
\angle**=0\clap{\vrule width#1\L* height.1pt}\else
\L*=#1\L*\L*=.5\L*\rmov*(-\L*,0pt){\sm*}\rmov*(\L*,0pt){\sl*}\fi}

\def\shade#1[#2]{\rlap{\Lengthunit=#1\Lengthunit
\special{em:linewidth .001pt}\relax
\mov(0,#2.05){\hr*{.994}}\mov(0,#2.1){\hr*{.980}}\relax
\mov(0,#2.15){\hr*{.953}}\mov(0,#2.2){\hr*{.916}}\relax
\mov(0,#2.25){\hr*{.867}}\mov(0,#2.3){\hr*{.798}}\relax
\mov(0,#2.35){\hr*{.715}}\mov(0,#2.4){\hr*{.603}}\relax
\mov(0,#2.45){\hr*{.435}}\special{em:linewidth \the\linwid*}}}

\def\dshade#1[#2]{\rlap{\special{em:linewidth .001pt}\relax
\Lengthunit=#1\Lengthunit\if#2-\def\t*{+}\else\def\t*{-}\fi
\mov(0,\t*.025){\relax
\mov(0,#2.05){\hr*{.995}}\mov(0,#2.1){\hr*{.988}}\relax
\mov(0,#2.15){\hr*{.969}}\mov(0,#2.2){\hr*{.937}}\relax
\mov(0,#2.25){\hr*{.893}}\mov(0,#2.3){\hr*{.836}}\relax
\mov(0,#2.35){\hr*{.760}}\mov(0,#2.4){\hr*{.662}}\relax
\mov(0,#2.45){\hr*{.531}}\mov(0,#2.5){\hr*{.320}}\relax
\special{em:linewidth \the\linwid*}}}}

\def\vdot{\rlap{\kern-1.9pt\lower1.8pt\hbox{$\scriptstyle\bullet$}}}
\def\vtimes{\rlap{\kern-3pt\lower1.8pt\hbox{$\scriptstyle\times$}}}
\def\vDot{\rlap{\kern-2.3pt\lower2.7pt\hbox{$\bullet$}}}
\def\vTimes{\rlap{\kern-3.6pt\lower2.4pt\hbox{$\times$}}}

\def\arc(#1)[#2,#3]{{\k*=#2\l*=#3\m*=\l*
\advance\m*-6\ifnum\k*>\l*\relax\else
{\rotate(#2)\mov(#1,0){\sm*}}\loop
\ifnum\k*<\m*\advance\k*5{\rotate(\k*)\mov(#1,0){\sl*}}\repeat
{\rotate(#3)\mov(#1,0){\sl*}}\fi}}

\def\dasharc(#1)[#2,#3]{{\k**=#2\n*=#3\advance\n*-1\advance\n*-\k**
\L*=1000sp\L*#1\L* \multiply\L*\n* \multiply\L*\Nhalfperiods
\divide\L*57\N*\L* \divide\N*2000\ifnum\N*=0\N*1\fi
\r*\n*  \divide\r*\N* \ifnum\r*<2\r*2\fi
\m**\r* \divide\m**2 \l**\r* \advance\l**-\m** \N*\n* \divide\N*\r*
\k**\r* \multiply\k**\N* \dn*\n* \advance\dn*-\k** \divide\dn*2\advance\dn*\*one
\r*\l** \divide\r*2\advance\dn*\r* \advance\N*-2\k**#2\relax
\ifnum\l**<6{\rotate(#2)\mov(#1,0){\sm*}}\advance\k**\dn*
{\rotate(\k**)\mov(#1,0){\sl*}}\advance\k**\m**
{\rotate(\k**)\mov(#1,0){\sm*}}\loop
\advance\k**\l**{\rotate(\k**)\mov(#1,0){\sl*}}\advance\k**\m**
{\rotate(\k**)\mov(#1,0){\sm*}}\advance\N*-1\ifnum\N*>0\repeat
{\rotate(#3)\mov(#1,0){\sl*}}\else\advance\k**\dn*
\arc(#1)[#2,\k**]\loop\advance\k**\m** \r*\k**
\advance\k**\l** {\arc(#1)[\r*,\k**]}\relax
\advance\N*-1\ifnum\N*>0\repeat
\advance\k**\m**\arc(#1)[\k**,#3]\fi}}

\def\triangarc#1(#2)[#3,#4]{{\k**=#3\n*=#4\advance\n*-\k**
\L*=1000sp\L*#2\L* \multiply\L*\n* \multiply\L*\Nhalfperiods
\divide\L*57\N*\L* \divide\N*1000\ifnum\N*=0\N*1\fi
\d**=#2\Lengthunit \d*\d** \divide\d*57\multiply\d*\n*
\r*\n*  \divide\r*\N* \ifnum\r*<2\r*2\fi
\m**\r* \divide\m**2 \l**\r* \advance\l**-\m** \N*\n* \divide\N*\r*
\dt*\d* \divide\dt*\N* \dt*.5\dt* \dt*#1\dt*
\divide\dt*1000\multiply\dt*\magnitude
\k**\r* \multiply\k**\N* \dn*\n* \advance\dn*-\k** \divide\dn*2\relax
\r*\l** \divide\r*2\advance\dn*\r* \advance\N*-1\k**#3\relax
{\rotate(#3)\mov(#2,0){\sm*}}\advance\k**\dn*
{\rotate(\k**)\mov(#2,0){\sl*}}\advance\k**-\m**\advance\l**\m**\loop\dt*-\dt*
\d*\d** \advance\d*\dt*
\advance\k**\l**{\rotate(\k**)\rmov*(\d*,0pt){\sl*}}%
\advance\N*-1\ifnum\N*>0\repeat\advance\k**\m**
{\rotate(\k**)\mov(#2,0){\sl*}}{\rotate(#4)\mov(#2,0){\sl*}}}}

\def\wavearc#1(#2)[#3,#4]{{\k**=#3\n*=#4\advance\n*-\k**
\L*=4000sp\L*#2\L* \multiply\L*\n* \multiply\L*\Nhalfperiods
\divide\L*57\N*\L* \divide\N*1000\ifnum\N*=0\N*1\fi
\d**=#2\Lengthunit \d*\d** \divide\d*57\multiply\d*\n*
\r*\n*  \divide\r*\N* \ifnum\r*=0\r*1\fi
\m**\r* \divide\m**2 \l**\r* \advance\l**-\m** \N*\n* \divide\N*\r*
\dt*\d* \divide\dt*\N* \dt*.7\dt* \dt*#1\dt*
\divide\dt*1000\multiply\dt*\magnitude
\k**\r* \multiply\k**\N* \dn*\n* \advance\dn*-\k** \divide\dn*2\relax
\divide\N*4\advance\N*-1\k**#3\relax
{\rotate(#3)\mov(#2,0){\sm*}}\advance\k**\dn*
{\rotate(\k**)\mov(#2,0){\sl*}}\advance\k**-\m**\advance\l**\m**\loop\dt*-\dt*
\d*\d** \advance\d*\dt* \dd*\d** \advance\dd*1.3\dt*
\advance\k**\r*{\rotate(\k**)\rmov*(\d*,0pt){\sl*}}\relax
\advance\k**\r*{\rotate(\k**)\rmov*(\dd*,0pt){\sl*}}\relax
\advance\k**\r*{\rotate(\k**)\rmov*(\d*,0pt){\sl*}}\relax
\advance\k**\r*
\advance\N*-1\ifnum\N*>0\repeat\advance\k**\m**
{\rotate(\k**)\mov(#2,0){\sl*}}{\rotate(#4)\mov(#2,0){\sl*}}}}

\def\gmov*#1(#2,#3)#4{\rlap{\L*=#1\Lengthunit
\xL*=#2\L* \yL*=#3\L*
\rx* \gcos*\xL* \tmp* \gsin*\yL* \advance\rx*-\tmp*
\ry* \gcos*\yL* \tmp* \gsin*\xL* \advance\ry*\tmp*
\rx*=\xscale\rx* \ry*=\yscale\ry*
\xL* \the\cos*\rx* \tmp* \the\sin*\ry* \advance\xL*-\tmp*
\yL* \the\cos*\ry* \tmp* \the\sin*\rx* \advance\yL*\tmp*
\kern\xL*\raise\yL*\hbox{#4}}}

\def\rgmov*(#1,#2)#3{\rlap{\xL*#1\yL*#2\relax
\rx* \gcos*\xL* \tmp* \gsin*\yL* \advance\rx*-\tmp*
\ry* \gcos*\yL* \tmp* \gsin*\xL* \advance\ry*\tmp*
\rx*=\xscale\rx* \ry*=\yscale\ry*
\xL* \the\cos*\rx* \tmp* \the\sin*\ry* \advance\xL*-\tmp*
\yL* \the\cos*\ry* \tmp* \the\sin*\rx* \advance\yL*\tmp*
\kern\xL*\raise\yL*\hbox{#3}}}

\def\Earc(#1)[#2,#3][#4,#5]{{\k*=#2\l*=#3\m*=\l*
\advance\m*-6\ifnum\k*>\l*\relax\else\def\xscale{#4}\def\yscale{#5}\relax
{\angle**0\rotate(#2)}\gmov*(#1,0){\sm*}\loop
\ifnum\k*<\m*\advance\k*5\relax
{\angle**0\rotate(\k*)}\gmov*(#1,0){\sl*}\repeat
{\angle**0\rotate(#3)}\gmov*(#1,0){\sl*}\relax
\def\xscale{1}\def\yscale{1}\fi}}

\def\dashEarc(#1)[#2,#3][#4,#5]{{\k**=#2\n*=#3\advance\n*-1\advance\n*-\k**
\L*=1000sp\L*#1\L* \multiply\L*\n* \multiply\L*\Nhalfperiods
\divide\L*57\N*\L* \divide\N*2000\ifnum\N*=0\N*1\fi
\r*\n*  \divide\r*\N* \ifnum\r*<2\r*2\fi
\m**\r* \divide\m**2 \l**\r* \advance\l**-\m** \N*\n* \divide\N*\r*
\k**\r*\multiply\k**\N* \dn*\n* \advance\dn*-\k** \divide\dn*2\advance\dn*\*one
\r*\l** \divide\r*2\advance\dn*\r* \advance\N*-2\k**#2\relax
\ifnum\l**<6\def\xscale{#4}\def\yscale{#5}\relax
{\angle**0\rotate(#2)}\gmov*(#1,0){\sm*}\advance\k**\dn*
{\angle**0\rotate(\k**)}\gmov*(#1,0){\sl*}\advance\k**\m**
{\angle**0\rotate(\k**)}\gmov*(#1,0){\sm*}\loop
\advance\k**\l**{\angle**0\rotate(\k**)}\gmov*(#1,0){\sl*}\advance\k**\m**
{\angle**0\rotate(\k**)}\gmov*(#1,0){\sm*}\advance\N*-1\ifnum\N*>0\repeat
{\angle**0\rotate(#3)}\gmov*(#1,0){\sl*}\def\xscale{1}\def\yscale{1}\else
\advance\k**\dn* \Earc(#1)[#2,\k**][#4,#5]\loop\advance\k**\m** \r*\k**
\advance\k**\l** {\Earc(#1)[\r*,\k**][#4,#5]}\relax
\advance\N*-1\ifnum\N*>0\repeat
\advance\k**\m**\Earc(#1)[\k**,#3][#4,#5]\fi}}

\def\triangEarc#1(#2)[#3,#4][#5,#6]{{\k**=#3\n*=#4\advance\n*-\k**
\L*=1000sp\L*#2\L* \multiply\L*\n* \multiply\L*\Nhalfperiods
\divide\L*57\N*\L* \divide\N*1000\ifnum\N*=0\N*1\fi
\d**=#2\Lengthunit \d*\d** \divide\d*57\multiply\d*\n*
\r*\n*  \divide\r*\N* \ifnum\r*<2\r*2\fi
\m**\r* \divide\m**2 \l**\r* \advance\l**-\m** \N*\n* \divide\N*\r*
\dt*\d* \divide\dt*\N* \dt*.5\dt* \dt*#1\dt*
\divide\dt*1000\multiply\dt*\magnitude
\k**\r* \multiply\k**\N* \dn*\n* \advance\dn*-\k** \divide\dn*2\relax
\r*\l** \divide\r*2\advance\dn*\r* \advance\N*-1\k**#3\relax
\def\xscale{#5}\def\yscale{#6}\relax
{\angle**0\rotate(#3)}\gmov*(#2,0){\sm*}\advance\k**\dn*
{\angle**0\rotate(\k**)}\gmov*(#2,0){\sl*}\advance\k**-\m**
\advance\l**\m**\loop\dt*-\dt* \d*\d** \advance\d*\dt*
\advance\k**\l**{\angle**0\rotate(\k**)}\rgmov*(\d*,0pt){\sl*}\relax
\advance\N*-1\ifnum\N*>0\repeat\advance\k**\m**
{\angle**0\rotate(\k**)}\gmov*(#2,0){\sl*}\relax
{\angle**0\rotate(#4)}\gmov*(#2,0){\sl*}\def\xscale{1}\def\yscale{1}}}

\def\waveEarc#1(#2)[#3,#4][#5,#6]{{\k**=#3\n*=#4\advance\n*-\k**
\L*=4000sp\L*#2\L* \multiply\L*\n* \multiply\L*\Nhalfperiods
\divide\L*57\N*\L* \divide\N*1000\ifnum\N*=0\N*1\fi
\d**=#2\Lengthunit \d*\d** \divide\d*57\multiply\d*\n*
\r*\n*  \divide\r*\N* \ifnum\r*=0\r*1\fi
\m**\r* \divide\m**2 \l**\r* \advance\l**-\m** \N*\n* \divide\N*\r*
\dt*\d* \divide\dt*\N* \dt*.7\dt* \dt*#1\dt*
\divide\dt*1000\multiply\dt*\magnitude
\k**\r* \multiply\k**\N* \dn*\n* \advance\dn*-\k** \divide\dn*2\relax
\divide\N*4\advance\N*-1\k**#3\def\xscale{#5}\def\yscale{#6}\relax
{\angle**0\rotate(#3)}\gmov*(#2,0){\sm*}\advance\k**\dn*
{\angle**0\rotate(\k**)}\gmov*(#2,0){\sl*}\advance\k**-\m**
\advance\l**\m**\loop\dt*-\dt*
\d*\d** \advance\d*\dt* \dd*\d** \advance\dd*1.3\dt*
\advance\k**\r*{\angle**0\rotate(\k**)}\rgmov*(\d*,0pt){\sl*}\relax
\advance\k**\r*{\angle**0\rotate(\k**)}\rgmov*(\dd*,0pt){\sl*}\relax
\advance\k**\r*{\angle**0\rotate(\k**)}\rgmov*(\d*,0pt){\sl*}\relax
\advance\k**\r*
\advance\N*-1\ifnum\N*>0\repeat\advance\k**\m**
{\angle**0\rotate(\k**)}\gmov*(#2,0){\sl*}\relax
{\angle**0\rotate(#4)}\gmov*(#2,0){\sl*}\def\xscale{1}\def\yscale{1}}}

\newcount\CatcodeOfAtSign
\CatcodeOfAtSign=\the\catcode`\@
\catcode`\@=11
\def\@arc#1[#2][#3]{\rlap{\Lengthunit=#1\Lengthunit
\sm*\l*arc(#2.1914,#3.0381)[#2][#3]\relax
\mov(#2.1914,#3.0381){\l*arc(#2.1622,#3.1084)[#2][#3]}\relax
\mov(#2.3536,#3.1465){\l*arc(#2.1084,#3.1622)[#2][#3]}\relax
\mov(#2.4619,#3.3086){\l*arc(#2.0381,#3.1914)[#2][#3]}}}

\def\dash@arc#1[#2][#3]{\rlap{\Lengthunit=#1\Lengthunit
\d*arc(#2.1914,#3.0381)[#2][#3]\relax
\mov(#2.1914,#3.0381){\d*arc(#2.1622,#3.1084)[#2][#3]}\relax
\mov(#2.3536,#3.1465){\d*arc(#2.1084,#3.1622)[#2][#3]}\relax
\mov(#2.4619,#3.3086){\d*arc(#2.0381,#3.1914)[#2][#3]}}}

\def\wave@arc#1[#2][#3]{\rlap{\Lengthunit=#1\Lengthunit
\w*lin(#2.1914,#3.0381)\relax
\mov(#2.1914,#3.0381){\w*lin(#2.1622,#3.1084)}\relax
\mov(#2.3536,#3.1465){\w*lin(#2.1084,#3.1622)}\relax
\mov(#2.4619,#3.3086){\w*lin(#2.0381,#3.1914)}}}

\def\bezier#1(#2,#3)(#4,#5)(#6,#7){\N*#1\l*\N* \advance\l*\*one
\d* #4\Lengthunit \advance\d* -#2\Lengthunit \multiply\d* \*two
\b* #6\Lengthunit \advance\b* -#2\Lengthunit
\advance\b*-\d* \divide\b*\N*
\d** #5\Lengthunit \advance\d** -#3\Lengthunit \multiply\d** \*two
\b** #7\Lengthunit \advance\b** -#3\Lengthunit
\advance\b** -\d** \divide\b**\N*
\mov(#2,#3){\sm*{\loop\ifnum\m*<\l*
\a*\m*\b* \advance\a*\d* \divide\a*\N* \multiply\a*\m*
\a**\m*\b** \advance\a**\d** \divide\a**\N* \multiply\a**\m*
\rmov*(\a*,\a**){\unhcopy\spl*}\advance\m*\*one\repeat}}}

\catcode`\*=12

\newcount\n@ast
\def\n@ast@#1{\n@ast0\relax\get@ast@#1\end}
\def\get@ast@#1{\ifx#1\end\let\next\relax\else
\ifx#1*\advance\n@ast1\fi\let\next\get@ast@\fi\next}

\newif\if@up \newif\if@dwn
\def\up@down@#1{\@upfalse\@dwnfalse
\if#1u\@uptrue\fi\if#1U\@uptrue\fi\if#1+\@uptrue\fi
\if#1d\@dwntrue\fi\if#1D\@dwntrue\fi\if#1-\@dwntrue\fi}

\def\halfcirc#1(#2)[#3]{{\Lengthunit=#2\Lengthunit\up@down@{#3}\relax
\if@up\mov(0,.5){\@arc[-][-]\@arc[+][-]}\fi
\if@dwn\mov(0,-.5){\@arc[-][+]\@arc[+][+]}\fi
\def\lft{\mov(0,.5){\@arc[-][-]}\mov(0,-.5){\@arc[-][+]}}\relax
\def\rght{\mov(0,.5){\@arc[+][-]}\mov(0,-.5){\@arc[+][+]}}\relax
\if#3l\lft\fi\if#3L\lft\fi\if#3r\rght\fi\if#3R\rght\fi
\n@ast@{#1}\relax
\ifnum\n@ast>0\if@up\shade[+]\fi\if@dwn\shade[-]\fi\fi
\ifnum\n@ast>1\if@up\dshade[+]\fi\if@dwn\dshade[-]\fi\fi}}

\def\halfdashcirc(#1)[#2]{{\Lengthunit=#1\Lengthunit\up@down@{#2}\relax
\if@up\mov(0,.5){\dash@arc[-][-]\dash@arc[+][-]}\fi
\if@dwn\mov(0,-.5){\dash@arc[-][+]\dash@arc[+][+]}\fi
\def\lft{\mov(0,.5){\dash@arc[-][-]}\mov(0,-.5){\dash@arc[-][+]}}\relax
\def\rght{\mov(0,.5){\dash@arc[+][-]}\mov(0,-.5){\dash@arc[+][+]}}\relax
\if#2l\lft\fi\if#2L\lft\fi\if#2r\rght\fi\if#2R\rght\fi}}

\def\halfwavecirc(#1)[#2]{{\Lengthunit=#1\Lengthunit\up@down@{#2}\relax
\if@up\mov(0,.5){\wave@arc[-][-]\wave@arc[+][-]}\fi
\if@dwn\mov(0,-.5){\wave@arc[-][+]\wave@arc[+][+]}\fi
\def\lft{\mov(0,.5){\wave@arc[-][-]}\mov(0,-.5){\wave@arc[-][+]}}\relax
\def\rght{\mov(0,.5){\wave@arc[+][-]}\mov(0,-.5){\wave@arc[+][+]}}\relax
\if#2l\lft\fi\if#2L\lft\fi\if#2r\rght\fi\if#2R\rght\fi}}

\catcode`\*=11

\def\Circle#1(#2){\halfcirc#1(#2)[u]\halfcirc#1(#2)[d]\n@ast@{#1}\relax
\ifnum\n@ast>0\L*=\xscale\Lengthunit
\ifnum\angle**=0\clap{\vrule width#2\L* height.1pt}\else
\L*=#2\L*\L*=.5\L*\special{em:linewidth .001pt}\relax
\rmov*(-\L*,0pt){\sm*}\rmov*(\L*,0pt){\sl*}\relax
\special{em:linewidth \the\linwid*}\fi\fi}

\catcode`\*=12

\def\wavecirc(#1){\halfwavecirc(#1)[u]\halfwavecirc(#1)[d]}

\def\dashcirc(#1){\halfdashcirc(#1)[u]\halfdashcirc(#1)[d]}

\def\xscale{1}
\def\yscale{1}

\def\Ellipse#1(#2)[#3,#4]{\def\xscale{#3}\def\yscale{#4}\relax
\Circle#1(#2)\def\xscale{1}\def\yscale{1}}

\def\dashEllipse(#1)[#2,#3]{\def\xscale{#2}\def\yscale{#3}\relax
\dashcirc(#1)\def\xscale{1}\def\yscale{1}}

\def\waveEllipse(#1)[#2,#3]{\def\xscale{#2}\def\yscale{#3}\relax
\wavecirc(#1)\def\xscale{1}\def\yscale{1}}

\def\halfEllipse#1(#2)[#3][#4,#5]{\def\xscale{#4}\def\yscale{#5}\relax
\halfcirc#1(#2)[#3]\def\xscale{1}\def\yscale{1}}

\def\halfdashEllipse(#1)[#2][#3,#4]{\def\xscale{#3}\def\yscale{#4}\relax
\halfdashcirc(#1)[#2]\def\xscale{1}\def\yscale{1}}

\def\halfwaveEllipse(#1)[#2][#3,#4]{\def\xscale{#3}\def\yscale{#4}\relax
\halfwavecirc(#1)[#2]\def\xscale{1}\def\yscale{1}}

\catcode`\@=\the\CatcodeOfAtSign

\begin{titlepage}
\thispagestyle{empty}

\begin{flushright}
OHSTPY-HEP-T-01-22\\

\end{flushright}

\begin{center}
{\Large\bf Two-loop  $\N=4$  Super Yang  Mills effective action 
 \\
\vspace{2 mm}
and  interaction between  D3-branes}
\end{center}
\vspace{3mm}

\begin{center}
{\large I.L. Buchbinder${}^{ {\rm a}}$, 
A.Yu. Petrov$^{a,b}$  
and
A.A. Tseytlin$^{c,}$\footnote
{Also at Imperial College,
London and 
Lebedev Physics Institute, Moscow}
}\\
\vspace{2mm}

${}^a$\footnotesize{ {\it Department of Theoretical Physics\\
Tomsk State Pedagogical University\\
Tomsk 634041, Russia}}

${}^b$\footnotesize{
{\it
Instituto de F\'{i}sica, Universidade de S\~{a}o Paulo\\
P.O. Box 66318, 05315-970, S\~{a}o Paulo, Brasil}} 

$^c$\footnotesize{ {\it Department of Physics\\
The Ohio State University  \\
Columbus, OH 43210-1106, USA}}
\end{center}
\vspace{5mm}

\begin{abstract}
\baselineskip=14pt
We  compute the leading low-energy term in the planar part
of the 2-loop contribution to the effective action of $\N=4$  
  SYM theory in 4 dimensions, assuming that the gauge group 
$SU(N+1)$ is broken to $SU(N) \times  U(1)$ by a constant  
scalar background $X$. While the leading 1-loop
correction  is the  familiar $c_1  F^4/|X|^4$  term,
the 2-loop expression starts with  $c_2 F^6/|X|^8$.
The 1-loop constant $c_1$ is known to be equal to   
the  coefficient of the  $F^4$ term in the
Born-Infeld action for a probe D3-brane separated
by  distance $|X|$ from a large number $N$  of  coincident
D3-branes. We show that the same is true  also for the
2-loop  constant $c_2$: it matches  the coefficient 
of the $F^6$ term in the D3-brane probe action.
In the context of the AdS/CFT correspondence, this  agreement
suggests a non-renormalization of the coefficient of the 
$F^6$ term beyond two loops. Thus the result of hep-th/9706072
about the agreement between the $v^6$ term in the D0-brane
supergravity interaction potential and the corresponding 
2-loop term in the 1+0 dimensional reduction of $\N=4$ 
SYM theory  has indeed a direct 
generalization to 1+3 dimensions,
as conjectured  earlier in hep-th/9709087. We also discuss
the issue of gauge theory -- supergravity  correspondence  for 
higher order ($F^8$, etc.) terms.

\end{abstract}
\vfill
\end{titlepage}


\newpage

\renewcommand{\thefootnote}{\arabic{footnote}}
\setcounter{footnote}{0}

\def \ff {{\rm  f}}
\def \P {\Phi} 

\section{Introduction}
The  remarkable relation   between
supersymmetric gauge theories
and  supergravity
implied by existence  of D-branes in string theory 
\ci{pol,ed,ig,mal}
motives  detailed study
of quantum  corrections in  super Yang-Mills theory,
and, in particular, their non-renormalization properties.
One  aspect  of this relation  which will be
of interest to us here  is a  correspondence
between the $\N=4$ super Yang-Mills
  theory and  type IIB supergravity descriptions
of subleading  terms in the  interaction potential
between parallel  D3-branes
(see, e.g., \ci{TAYL,doug,banks,Lif,Mald,ch2,bbpt,ch3}
for related discussions of interactions between D-branes).


Consider the  supergravity-implied
 action for a D3-brane probe  in curved
background produced by a large  number $N$  of
coincident D3-branes.
Ignoring higher-derivative (``acceleration")
terms,  it  is given by the
 Born-Infeld action in the corresponding curved
  metric and the ``electric'' part of the  R-R 4-form potential,
  \be
S =  - T_3  \int d^{4}  x \  H^{-1} (X)
     \bigg[ \sqrt {-\det (  \eta_{mn}  +
H(X)  \del_m X^i \del_n X^i  +
 H^{1/2} (X)  F_{mn} )} -1\bigg]    \ .
\la{prob}
\ee
Here $i=1,...,6$,\ $m,n=0,1,2,3$, \
  $T_3 = {1 \ov  2 \pi g_s }$ 
  and  $H= 1 + {Q\ov |X|^4} $, \ $
  Q\equiv { 1 \ov \pi}   N g_s $.\foot{We set string tension
  $T= {1 \ov 2 \pi \a'}$ to be 1. In general,
  $T^{-1}$ appears  in front of $F_{mn}$ and
  in the relation between
  the scale $X$ in the supergravity expressions and the scalar
  expectation
  value $\Phi$ in the SYM expressions.}
In addition, the action contains also the 
``magnetic" interaction part given
by the Chern-Simons term,
 $ S_{\rm mag}= N S_{\rm WZ}\sim  i N \int_{5}
 \epsilon_{i_1  ...i_6} { 1 \ov |X|^6}   X^{i_1}
  d
X^{i_2} \wedge ...\wedge d X^{i_6}$.
In what follows we shall consider the case when
 $X^i=\const$, i.e. ignore all scalar derivative terms.
  We shall  assume that
${Q\ov |X|^4}  \gg 1$, i.e. that one can  drop 1 in
the harmonic function $H$,  so that \rf{prob}
becomes the same as  the action for a D3-brane probe in 
the 
 $AdS_5 \times S^5$ space (oriented parallel to the  boundary
 of $AdS_5$)
   \be
S =  - T_3  \int d^{4}  x \  {|X|^4\ov Q}
     \bigg[ \sqrt {-\det (  \eta_{mn}  +
{ Q^{1/2} \ov |X|^2 }   F_{mn} )} -1\bigg]  \ .
\la{prb}
\ee
Expanding in powers of $F$, we get 
\be
S=  - T_3  \int d^{4}  x\biggl(  -  { 1\ov 4} F^2  - { 1\ov
8}
{Q\ov |X|^4}
 \big[F^4 - { 1\ov 4} (F^2)^2 \big]
- { 1\ov 12}  \big({Q\ov |X|^4}\big)^2
 \big[F^6 - { 3 \ov 8 } F^4 F^2 + { 1 \ov 32} ( F^2)^3\big]
 + ... \biggr)
 \ ,
\la{see}
\ee
where $F^k$ is the trace of the matrix product in Lorentz
indices,
i.e.
$  F^2 = F_{mn} F_{nm}= - F^2_{mn} , \  \  F^k = F_{m_1m_2}
 F_{m_2m_3} ... F_{m_km_1}.$\foot{Note that the sum of
 the $F^4$ and $F^6$  can be represented as follows:
 $  - { 1\ov 8}
{Q\ov |X|^4}
 \big[F^4 - { 1\ov 4} (F^2)^2 \big] ( 1 - 
 { 1\ov 4} {Q\ov |X|^4} F_{mn} F_{mn})
   $
 (in fact, all higher terms in the expansion of
 the $D=4$ BI action  are proportional to the  $F^4$ term
  \ci{TR}). }
 The general structure of this expansion is thus
\be
S= {1 \ov g_s}\int d^{4}  x  \ \sum^\infty_{l=0}\  {\rm c}_l\  (g_s N)^{l}
{ F^{2l+2}\over |X|^{4l}} \ .
\label{form}
\ee
{}From the weakly-coupled  flat-space
 string theory point of view, the leading-order
interactions between  separated D-branes  are 
described by  the ``disc with holes''  diagrams
(i.e. annulus, etc).
The limit of small separation should be 
represented  by loop
corrections in SYM theory, while  the limit of large
separation -- by classical supergravity exchanges.
If the coefficient of a particular term in the string
interaction potential (like  $v^4$ term in
\ci{doug})
happens not to depend on the distance
(i.e. on dimensionless ratio  of
separation and $\sqrt{\a'}$)   then its coefficient
should be the same
in the quantum SYM and the  classical supergravity expressions  for
 the interaction.

In  the SYM  theory  language, computing the 
interaction potential between a stack of D3-branes 
and a parallel D3-brane probe carrying 
 constant $F_{mn}$ background field  
 corresponds to computing
the quantum effective action $\G$ 
in   constant  scalar $\P^i $ background
which breaks $SU(N+1)$ to $SU(N)\times U(1)$
and  in constant
$U(1)$ gauge field $F_{mn}$.
 For the interactions between {\it  D3}-branes,
i.e.  in   the case of  finite  $\N=4, \ D=4$ SYM  theory,
  the
expansion
of $\G$ in  powers of the dimensionless ratio $F^2/|\P|^4$
has   the  following general form
\be
\G= { 1 \ov  \gym } \int d^{4}  x  \ \sum^\infty_{l=0}
f_l ( \gym,  N)
{ F^{2l+2}\over |\P|^{4l}} \ .
\label{yorm}
\ee
In  the planar  (large $N$, fixed  $\l\equiv  \gym  N$)
approximation  the functions $f_l$
should depend only on $\l$.

In more detail,  the planar $l$-loop  diagrams  we
are interested in are the ones where the  background field legs 
are attached to the ``outer" boundary only (in double-line notation).  
In D-brane interaction picture, this corresponds to 
  one  boundary of the $l$-loop graph  attached to 
  one D-brane (which carries $F_{mn}$
background) and  all other $l$  boundaries  attached to  $N$ coincident
``empty'' D-branes. This produces  the factor of $ N^{l}$.

The comparison of \rf{yorm}
with the supergravity expression \rf{form}
is done for $\gym= 2 \pi g_s$  and $|\P|= T |X| $. 
Naively, it  would work   term-by-term if 
$f_l ( \gym N) = c_l (\gym N)^{l} $,
i.e. if the  $l$-th term in \rf{yorm} would  receive
 contribution
{\it only} from the $l$-th loop order.

This is indeed   what is known to happen for  the leading
$F^4/|X|^4$ term\foot{Since we
set $T=1$ (see footnote 1)
in what follows we shall not distinguish between $\P$ and $X$.}
 which  appears
only at the first loop order, and not at higher orders
due to the existence of a   non-renormalization
theorem \ci{DS}.\foot{The absence of the 2-loop
$F^4$ correction was proved 
 in 1+3 dimensional $\N=4$ SYM  theory in \ci{bko},
 and similar result  was found in  1+0  dimensional theory
 \ci{BB}. For  indications of existence
 of more general
 non-renormalization theorems see \ci{dig}.
 }
 Moreover, the  one-loop
coefficient  of the $F^4$ term in the
 $\N=4$ SYM effective action
\ci{GS,FT} is    in
precise agreement  with the supergravity
expression (see, e.g., \ci{TAYL,ch3,Mald}).

In the D3-brane case, there is also another
 viewpoint suggesting a  correspondence
 between  the classical  supergravity D3-brane probe action
 \rf{prb}   and the quantum 
 SYM  effective action \rf{yorm} --
 the AdS/CFT conjecture \ci{mal,other}.
 In the present context it
  implies that the supergravity action  \rf{prb}
 should  agree with  the strong `t Hooft coupling 
 limit of the planar (large $N$) part 
 of the SYM  action \rf{yorm}.
 The AdS/CFT conjecture thus imposes  a weaker
 restriction that $f_l (\l)_{\l \gg 1} \to a_l \l^l$,
 with $a_l$ being  directly related  to
   ${\rm c}_l$ in  \rf{prb}.
  The simplest possibility to satisfy this 
  condition  would be  realized if 
    the  functions of coupling in front 
     of some of the
  terms  in  \rf{yorm}  would
    receive 
  contributions only from the 
  particular orders in perturbation theory,  and  with
   the right coefficients to match \rf{prb},
 i.e. if they 
 would  not be  renormalized  by all {\it higher}-loop 
    corrections. As we shall see, while this is likely to 
    be the case  at the $F^6$ order, 
    the situation for $F^8$, etc.,  terms  is bound to  be 
     more complicated.

\subsection{ The $F^6$ term}

In  this paper  we shall study
this  correspondence  for
the $F^6$ term  by explicitly
computing its 2-loop coefficient  in the $\N=4$ SYM theory.
The Lorentz structure of this term in the 
SYM effective action 
is the same as in the BI action \rf{see}
(the form of the abelian $F^6$ term  is, in fact, 
  fixed uniquely by the $\N=1$ supersymmetry), 
and  the  planar ($N\gg 1$)  part of  its coefficient
 will turn out to be
  exactly the {same}  as appearing in \rf{form}.

This precise agreement between the supergravity and the SYM
actions at the
 $F^6$  level we shall  establish  below 
was  conjectured earlier in \ci{ch3},\foot{It was
further  conjectured in \ci{ch3}
that all terms in the BI action may be reproduced
from the SYM effective action (see also \ci{Mald,mal,kesk}
for  similar conjectures). We shall present a  more 
precise version of this conjecture  in section 1.2 and 
section 5. }
  being
 motivated  by the  agreement
\ci{bbpt}  between  the $v^6/|X|^{14}$  term in the
interaction
potential between D0-branes  in the  supergravity
description and  the  corresponding
2-loop term  in the effective action
of maximally supersymmetric 1+0 dimensional SYM
theory.\foot{This
conjecture was tested in \ci{ch3}
 (in the general non-abelian case)
by  demonstrating   that
there is a universal   $N\gym {F^6\ov |X|^8} $ expression
  on the SYM side  which
reproduces  subleading terms in  the  supergravity  potentials
between  various bound-state  configurations of branes.
Since
brane systems with different amounts of supersymmetry
are described by  very different SYM backgrounds,
the  assumption  that
all of the corresponding interaction potentials originate
from  a single universal   SYM  expression
provided highly non-trivial constraints on the
structure of the  latter.}

Combined with the known fact that the abelian $F^6$ term
does  not appear at  the {\it one-loop} $\N=4$ 
SYM effective action 
\ci{FT,ch2} (see eq.\rf{yumi} below),
this  suggests that this
2-loop coefficient should be  exact, i.e.  the abelian
$F^6$  term should 
not receive contributions from all higher ($l\geq 3$) 
loop orders.
Indeed, from the point  of  view of the
AdS/CFT correspondence,
 higher order $(\gym N)^n$ corrections to $f_2$ in \rf{yorm} 
 would
dominate over the two-loop one for $\l \gg 1$,
 spoiling the interpretation
of \rf{form} as the strong-coupling limit of \rf{yorm}.

One  should thus expect 
 the existence of a new non-renormalization
theorem
for the abelian $F^6$ term in $\G$ (computed in the  planar
approximation),  analogous  
to the well-known $F^4$ theorem  of   \ci{DS}.
The reasoning used in \ci{DS} (based on scale invariance and
$\N=2$
supersymmetry)
does not, however, seem to be  enough to prove the non-renormalization
of the $F^6$ term. It is most likely that one needs to use 
 the
full power  of 16 supersymmetries of the theory
(which are  realized in a ``deformed"  way).
One may then expect to show that the $\N=4$ supersymmetry demands
that the coefficient of the $F^6$  term should be rigidly fixed
in terms of the $F^4$-coefficient (proportional to 
its square); then  the fact the $F^4$ term appears only 
at the  1-loop order would 
imply that the $F^6$ term should be present  
 only at the 2-loop order. 

One  possible way to  demonstrate this would  be 
to   apply  the component approach, 
by generalizing to 1+3 dimensions
what   was done for the $v^6/|X|^{14}$ term
 in 1+0 dimensions  \ci{paban} (see also \ci{kaz}),
i.e. by  deforming the
supersymmetry  transformation rules
order by order in $1/|X|^2$  and trying  to  show  that
the coefficient of the $F^6/|X|^{8}$  term in $\G$
is completely fixed by the 
supersymmetry in terms of  the
coefficient of the  $F^4/|X|^{4}$ term.
In effect, this is what was already done in  \ci{mrt}
in $D=10, \N=1$ SYM theory in a $U(1)$ background.
 It was shown there  that the structure of  the abelian
$F^4$ and $F^6$  terms in the
effective action  which starts with the  super Maxwell term
 is completely  fixed by the (deformed) 
  $\N=1, D=10$ supersymmetry
to be the same as in the BI action,\foot{In $D=10$
the role of $ 1/|X|$ or a  fundamental scale
is played by an UV cutoff  (or $\sqrt{\a'}$), 
powers of which multiply $F^n$.}
 with  the coefficient
of the $F^6$ term being  related to that  of   the $F^4$
term in
precisely the same  way as it comes out of 
 the expansion of the BI action.\foot{Let us mention also 
 that the coefficient of the $F^6$ term is  fixed  uniquely 
 in terms of the coefficient of that of  $F^4$ term 
 (to be exactly as in the  BI action)  by the condition of 
 self-duality of the $\N=4$  SYM effective action written in terms of 
 $\N=2$ superfields \ci{gonz,kut}.} 
We thus  expect that 
the arguments of \ci{mrt,paban} may indeed 
have a direct counterpart in $D=4$ theory, 
 relating the $F^4$ and $F^6$ coefficients  and 
and thus 
 providing  a proof  of 
the non-renormalization of  the $F^6$
 term beyond the 2-loop order.

\subsection{Comments on higher-order terms}
The  obvious  question  then is 
 what happens at the next --
 $F^8$  order:
should one  expect  that the coefficients
of these terms in the SYM effective action are 
again receiving contributions only from the corresponding --
3-loop
-- graphs  and that they are 
 in agreement with the supergravity expression
\rf{prb}?
That the story for the $F^8$ term  should be 
 {\it different}
from  the one for the  $F^6$ (and $F^4$) term 
 is indicated 
by the fact  that
 the 1-loop SYM effective action  \ci{FT,ch2}
 is already containing   a non-trivial 
  $F^8$ term (see \rf{yumi}).
 One  could still    hope
that the $F^8$ term will not receive corrections beyond
the 3-loop order,  so that the 3-loop contribution  will dominate
over the  1-loop and 2-loop  ones  
in the supergravity limit ($N\gym  \gg 1, \ N \gg 1$).

However, the situation is more complicated 
since,  in contrast to what was the case for the 
 $F^4$ and $F^6$ terms, the 
supersymmetry alone does {\it not}   constrain
the structure of the $F^8$ invariant
in a  unique way --
the $F^8$ terms in the BI action and in the 1-loop SYM
effective action
have, in fact,
  {\it different} Lorentz index structures!

Indeed,  considering four  Euclidean dimensions  and
   choosing the $U(1)$ gauge field background
    $F_{mn}$ to have  canonical
block-diagonal form with non-zero entries
$\ff_1=F_{12}$ and $\ff_2=F_{34} $
one  can  explicitly compare
 the  expansions of the  BI  action  and
 the 1-loop  SYM effective action.
 For the BI action we get 
 (we use a  constant $s$ instead of $H^{1/2}$ in \rf{prb}
 to  facilitate   comparison with  the SYM expression)
$$
\sqrt { \det ( \delta_{mn} +  s F_{mn} ) }
= 1 + \hf  (\ff^2_1 + \ff^2_2) s^2-
{ 1 \ov 8}  (\ff^2_1 - \ff^2_2)^2 s^4
$$
\be
+ \ { 1 \ov 16}
 (\ff^2_1 - \ff^2_2)^2(\ff^2_1 + \ff^2_2)s^6
 - { 1 \ov 128}  (\ff^2_1 - \ff^2_2)^2 ( 5 \ff^4_1 + 6
 \ff^2_1  \ff^2_2
+ 5 \ff^4_2) s^8  + O(s^{10} ) \ . \la{bii}
\ee
 The   1-loop  Euclidean Schwinger-type  effective  action
for  the $\N=4$     SYM theory  (with gauge group $SU(N+1)$ broken
to $SU(N) \times U(1)$  by a scalar field background $X$) 
depending on a constant  $U(1)$ 
gauge field  strength 
 $F_{mn}$ parametrized by $(\ff_1 , \ff_2)$
  has the following form \ci{FT,ch2}
\foot{We shall  
consider the $U(1)$ background  representing 
 a single D3-brane with gauge field ${ F}_{mn}$ 
separated (in one of the 6 transverse directions)
 by a  distance $X$ 
from  $N$ coincident D3-branes.
The  $U(N+1)$ background  matrix   in the  fundamental 
representation  is then 
   $\hat  F_{mn} =\diag( F_{mn}, 0, ..., 0)$ and the corresponding 
  $SU(N+1)$ matrix is $ F =  \hat F  - { \tr \hat F \ov N+1} I $, 
where  $\tr$  is  the trace in the  fundamental representation.
Only this traceless part of the $U(N+1)$ background couples to quantum fields
and  thus enters the quantum effective action 
(the planar part of which will  be proportional to a single trace 
of $F$ in the fundamental representaion). 
The (Minkowski space) YM
Lagrangian  will be    
$
{L_{\rm YM}} = - \frac{N}{4\l} {\rm tr} (F_{mn} F_{mn})  ,
$
where $\l = g_{\rm YM}^2 N$ and  the 
 generators in the fundamental representation
 are normalized  so that ${\rm tr} (T_I T_J) =  \delta_{IJ}$.
 This normalization is convenient  for comparison  with the kinetic 
 term in   the 
 BI action  for a collection of D3-branes
 (${L_{\rm YM}}$ should be evaluated 
 on the diagonal background  $U(N+1)$ 
 matrix in the fundamental representation).
In this case  $\l = 2\pi g_s N , \ \gym =  2\pi g_s.$
Comparing to the supergravity expression \rf{prb}  note that
for $2\pi \a'=1$ one has 
 $Q=   \frac{2 N\gym } {(2\pi)^2} $;
 note also that going from Euclidean to Minkowski signature 
 notation  one should change
 the overall sign of the action.
}
\begin{equation}
 \Gamma^{(1)}_E  =- {4 N V_4\ov (4\pi)^2}
\int\limits_{0}^{\infty }\frac{ds}{s^3}\, e^{ - s |X|^2} \
K(s)\ ,
\ \la{exx}
\ee
$$
K(s) =
{ \ff_1 s \ov \sinh \ff_1 s}\  { \ff_2 s \ov \sinh \ff_2 s}\
(\cosh  \ff_1 s - \cosh \ff_2 s)^2
$$
\be
= \ { 1 \ov 4} (\ff^2_1 - \ff^2_2)^2 s^4  + 0 \times s^6
+ { 1 \ov 960} (\ff^2_1 - \ff^2_2)^4  s^8 + O(s^{10} )\ .
\la{yumi}
\ee
The $F^6$ term in \rf{yumi}  cancels out,
but $F^8\sim \ff^8$ term is present and  has 
the structure different from that of the $F^8$ term 
in \rf{bii}.

Let  $(F^8)_1$ and $(F^8)_2$ denote the 
bosonic parts of the two
abelian super-invariants  appearing
at the $F^8$ order   in  the expansions of the BI
 action  \rf{bii} and   the  SYM  1-loop action
\rf{yumi} respectively.
We propose the following {\it conjecture}
  about the SYM effective
action
(which replaces the   earlier  conjecture in \ci{ch3}):
(i) the coefficient of $(F^8)_1$ receives contribution only
from the {\it 3-loop} order  with coefficient  which is
in precise agreement with the  one in the  supergravity BI
action
\rf{prb}; (ii) the coefficient of $(F^8)_2$ receives
contributions from {\it all}  loop orders, but the planar part
of  the resulting  non-trivial
function  $f_3 (N\gym )$ in \rf{yorm}   goes to zero
in the limit $N\gym  \gg 1$, so that the requirement
of the AdS/CFT correspondence is satisfied.

The possible  existence of the {\it  two} independent 
 invariants -- one of which
has  ``protected'' coefficient and another does not  
is reminiscent of 
 the situation for the $R^4$ invariants in type IIA string
theory
(see   \ci{TTT} and refs. there).
While a ``universality'' or
 ``BPS saturation'' of  coefficients of terms
 with
higher than 6 powers of $F$ may seem less plausible,
there are, in fact,
string-theory examples  of {\it specific}
  higher-order terms that receive
contributions
only from one particular  loop order, to all orders
 in loop expansion \ci{ant}.

 The existence of several independent  
 super-invariants  appearing at the same 
  $F^n$-order of  low-energy expansion of SYM effective action 
  with  only {\it one}  invariant  
 having  ``protected"  coefficient and 
 matching  onto the term appearing in the expansion
 of the BI action  may  then be  a general pattern.
 Moreover, it should probably apply  also in the  case of 
 {\it non-abelian}
  backgrounds, here 
   starting already  at  the $F^6$ order.

  Indeed, the non-abelian (or multi-$U(1)$)
  $F^6$ term is likely to be a combination
  of the two superinvariants  differing  by
   $[F,F]$ ``commutator'' terms   absent  in the   abelian limit.
  It is natural to expect  that there is, in fact,
  such commutator   term in the 1-loop SYM effective
  action.\foot{The
  precise structure and coefficient
   of this  term can be determined using  the general methods
   of  \ci{vand,shif,Delb},
  or by computing the non-abelian 
  $O(X^{12})$  1-loop term in the 1+0
 dimensional theory
  \ci{kabt}
  and lifting it to 1+9 or 1+3  dimensions.}
   Thus 
   the full non-abelian $F^6$ term in the SYM
   effective action will no longer  be protected. This
     may be related to  an   observation in 
     \ci{stern}
    that supersymmetry  does not 
  completely
   determine the coefficient of the  $O(v^6)$
   term in 1+0 dimensions  in the case of
   more general  (``$N >3$-particle")   $SU(N)$
    backgrounds.\foot{A possibility of existence, 
     in $SU(N)$, $N > 2$ case, 
   of ``unprotected"
   non-abelian 
   tensor structures already at ``$v^4$'' order
   (and that the proof of the  non-renormalization theorem of
   \ci{DS} applies in the 
    $SU(2)$  or  ``two brane" case only)
   was  suggested   in \ci{lowe}.}
   Particular diagonal $SU(3)$
     (``three D0-brane") backgrounds
   in 1+0 dimensional 2-loop SYM  effective action were
   considered in
   \ci{dira,okaw,zan} and the  agreement with supergravity
   was   demonstrated.

\subsection{Superspace form of the $F^6$ term}

Before describing the content of the technical part of the 
paper let us discuss the superspace form of the $F^6$ term in which 
it will be computed below.

 Using   $\N=1, D=4$ superfield notation,  the  expansion of the
  BI
 action  containing the
 sum of the $F^4$ and $F^6$ terms in 
   \rf{see}
  can be written as (cf. \ci{ferr}) 
$$
S =
  { 1 \ov  4 g^2_{\rm YM} }
  \bigg[\ ( \int {d}^6z \, W^2 + h.c.)  $$
  \be
+ \ \hf \  { 2 N g^2_{\rm YM} \over (2\pi)^2  |X|^4} \,  \int
{d}^8z \,
 {W^2\,{\bar W}^2  }
\bigg( 1  - { 1 \ov 16}   {  2 N g^2_{\rm YM} \over (2\pi)^2  |X|^4}
 (D^2\,W^2    + h.c.)
+ ...\bigg)\bigg]\  ,
 \la{www}
 \ee
where  $W$ is the abelian
$\N=1$ superfield strength. We assume  Minkowski signature choice 
as in  \rf{see}
and  the same  superspace conventions  as in \ci{bkt,kut} (in particular, 
$D^2 W^2|_{\theta=0}  = 2 F^2_{mn} + ...$).


To reproduce this expression   by 
a 2-loop  computation  on the   $\N=4$   SYM side
we shall use
the $\N=2$ superfield formulation
 (with the harmonic superspace
description for the quantum fields in the context of 
background field method).
 We shall  assume that
 only one $\N=1$ chiral superfield
 has a non-zero 
 constant expectation value, namely, the one contained
  in the $\N=2$ vector superfield, 
  i.e. only the latter and not the hypermultiplet 
   will have a non-vanishing background value.
Thus we will be 
 interested in  the case  when  the gauge group $SU(N+1)$
(or $U(N+1)$, which is the same  as we will consider only 
the leading large $N$ approximation)
is spontaneously broken down to $SU(N)\times U(1)$ by  the 
abelian constant  
(in $x$-space)      $\N=2$  superfield background   
  with non-zero components being  (same as in
 \ci{bkt}) 
  \be
 {\cW}|_{\q =0} = X = {\rm const}~,\qquad
\cD^i_{( \a} \cD_{\b)\, i} {\cW}|_{\q =0} = 8\,F_{\a \b}
= {\rm const}\ . \label{backg} \ee
It is understood that the   background 
function $  {\cW}$ 
 should be multiplied by 
 the diagonal $su(N+1)$ matrix (generating  the 
relevant abelian subgroup of $SU(N+1)$) 
 \be
 J = {\rm diag} (1, 0, ...,0) - { 1 \ov N +1 } I= 
{ 1 \ov N +1 }\diag(N, -1, ..., -1) \ , 
\la{beck} \ee
     which    
represents the configuration of  $N$ coincident 
D3-branes separated by a distance $X$ 
 from  the single  D3-brane carrying 
 the background $F_{mn}$ field.

Assuming  the  $\N=2$  background \rf{backg}, the  combination
of the  superconformal invariants
representing  the  sum of the two unique abelian 
$F^4/|X|^4$  and $F^6/|X|^8$   corrections in  \rf{see},\rf{www}
may be  written in the manifestly 
 $\N=2$ supersymmetric form  as follows 
\foot{We use that
for the given background  
$   \int
{d}^8z \,  { 1 \ov |X|^4}  
 {W^2\,{\bar W}^2  } =  
   \int d^{12}z     \ln\frac{\cal W}{\mu}
\ln\frac{\bar {\cal W}}{\mu} $ and 
$ \int d^8 z\frac{1}{|X|^8}W^2\bar{W}^2 D^2 W^2+ h.c.
= -\frac{1}{24}\int d^{12}z
\frac{1}{\bar{\cal W}^2}\ln \frac{\cal W}{\mu}\cD^4\ln \frac{\cal
W}{\mu} + h.c.$, etc.  These two $\N=2$ structures were discussed in, e.g., 
 \ci{dewi,DS} and \ci{gonz,bkt}, respectively.}
\begin{eqnarray}
S=& & \frac{1}{8\gym}(\int d^8 z\  {\cal W}^2  + h.c.) 
 \nonumber\\  &+& 
N \int d^{12} z\ \Big[ c_1 \ln\frac{\cal W}{\mu}
\ln\frac{\bar {\cal
W}}{\mu}+
 c_2  N\gym (\frac{1}{\bar{\cal W}^2} 
\ln\frac{\cal W}{\mu} {\cal D}^4\ln\frac{\cal W}{\mu}+h.c.)
\Big]  + ... \ , \la{twoo}
\end{eqnarray}
where (for the  relation between the $\N=1$ \rf{www}
and $\N=2$ \rf{twoo} forms of 
the $F^6$ term in
Appendix A)
\be \la{coee} c_1 = \frac{1}{ 4(2\pi)^2}\ , \ \ \ \ \ \ \ \ \ \ 
c_2 = \frac{1}{ 3\cdot 2^8 (2\pi)^4}   \ . \ee 
$\mu$ is a spurious scale  which  drops out
after one integrates  over $\theta$'s --
 it  gets replaced by 
${\cW}|_{\theta=0} = X $ in going from the 
$\N=2$ \rf{twoo}
to $\N=1$ \rf{www} form.\foot{For comparison, the $\N=2$ superfield form of
the
two abelian $F^8$ invariants  discussed above is  \ci{bkt,kut}:
$\int d^{12}z [
 \bar{\cW}^{-4} \ln\frac{\cW}{\m}  ({\cD^4}\ln \frac{\cW}{\m})^2  + h.c.)]
$ and $\int d^{12}z \bar{\cW}^{-2}{\cW}^{-2}
{\bar{\cD}^4}\ln\frac{\bar{\cW}}{\m}{\cD^4}
\frac{\cW}{\m}$, \ \   $d^{12} z = d^4 x d^4 \theta d^4 \bar \theta  $. }

The low-energy expansion of the quantum $\N=4$ SYM 
effective action  turns out to have the same structure \rf{twoo}.
The  coefficient of the 1-loop term $ \ln\frac{\cal W}{\mu}
\ln\frac{\bar {\cal W}}{\mu}$ 
 (computed  directly in  $\N=2$ 
 superspace form in \cite{roc1,bk1,gonz,lowe,bbk}) 
 coincides indeed with the corresponding coefficient $c_1$ 
 \rf{coee} in \rf{www}.\foot{The expression for the 1-loop 
(Minkowski-space)  effective action found in \cite{gonz,lowe,bbk}
 was \\ 
 $\G^{(1)}= { 1 \ov 16 \pi^2}  \sum_{k<l} 
 \int d^{12} z\ \ln\frac{ { \cal W}_k -{ \cal W}_l }{\mu}
\ln\frac{\bar {\cal W}_k -   \bar {\cal W}_l  }{\mu}, $\ 
where ${ \cal W}_k$ ($k=1,...,N+1$) are diagonal 
values of the background matrix in the fundamental representation of $su(N+1)$.
In the  case of the present  background \rf{backg},\rf{beck} 
the sum produces the factor of $N$ which matches the one in \rf{twoo}.}

Our aim  will  be to show that 
the  two-loop correction  to  the $\N=4$ SYM 
effective action in 
${\cW},\bar{\cW}$ background 
has the $\cN=2$ superspace form as the   $
\int d^{12} z\   (\frac{1}{\bar{\cal W}^2} 
\ln\frac{\cal W}{\mu} {\cal D}^4\ln\frac{\cal W}{\mu}+h.c.)$ 
    term   in 
\rf{twoo} with  the same coefficient  
$N^2 \gym c_2$ (in  the  large $N$ limit).
The  resulting conclusion will be that 
 both $F^4$ and $F^6$ terms in the SYM effective action  
coincide exactly with the terms in the 
second line of \rf{www}, i.e. with  the terms in the   BI action
in the  supergravity background.


\bigskip

The rest of the paper is organized as follows.
In section 2 we shall consider the 
${\cal N}=4$ super Yang-Mills theory formulated in
terms of unconstrained ${\cal N}=2$ superfields
 in harmonic superspace
\cite{GIKOS}, i.e. represented as the   ${\cal
N}=2$ SYM theory
coupled to hypermultiplet.
 Then we shall  briefly describe the ${\cal N}=2$ 
 superfield background
field method  allowing one 
to carry out the calculation of the  effective
action in a way 
preserving manifest ${\cal N}=2$ supersymmetry and gauge 
symmetry. Section 3
will be  devoted to the evaluation of the  hypermultiplet
 and ghost corrections.
  We shall find that their
 contributions to the  leading part of the  2-loop
 low-energy 
 effective action vanish.
  In section 
4 we shall compute  the  pure ${\cal N}=2$
SYM contribution to the 2-loop 
 $\N=4$ SYM effective action. 
Section 5 will contain a summary and 
some concluding remarks on  
possible generalizations.
Appendix A will present a
 relation between $\N=1$ and $\N=2$ forms of 
 the $F^6$. 
Appendix B  will describe some details 
 of  calculations of integrals over
 the  harmonic superspace.

\section{${\cal N}=2$  background superfield expansion
}
\setcounter{equation}{0}

The aim of this  work is to calculate
 the leading two-loop correction 
to the low-energy 
${\cal N}=4$ SYM effective action in the sector 
where  only the ${\cal N}=2$ vector
multiplet has a non-trivial background.
 Our starting point will be the  formulation of ${\cal N}=4$ SYM  theory
    in ${\cal N}=2$ harmonic superspace \cite{GIKOS}. 
 In  terms of $\N=2$ superfields 
${\cal N}=4$ SYM  is simply  ${\cal N}=2$ SYM theory interacting with
one  adjoint 
hypermultiplet with the  action  
\cite{bbko,bko}
\bea
\label{sq}
S_{{\N}=4 \ {\rm SYM}}=\frac{1}{4 g^2_{\rm YM}}{\rm tr}[\int d^8 z\  {\cal W}^2-
\int d\xi^{(-4)}\ \breve{q}^{+i}\nabla^{++}q^{+}_i] \ . 
\eea
 ${\cal W}$ is ${\cN}=2$ superfield strength expressed
in terms of ${\cN}=2$ vector  superfield
$V^{++}$, and
$q^{+i}=(\stackrel{\smile}{q}^+,-q^+)$ 
is the hypermultiplet field  taking values in
the  adjoint representation
of  gauge  algebra, with $q^+_i=(q^+,\stackrel{\smile}{q}^+)$.
Here and below  we use the notation  introduced in
 \cite{GIKOS,bbko,bko}.


The most natural  way to calculate  loop 
corrections in this  model  is 
to  use the  ${\cal N}=2$ background field method which guarantees 
manifest ${\cal N}=2$ supersymmetry and gauge covariance at each
 step of the 
calculation.
We shall do 
background-quantum splitting of the gauge superfield 
$V^{++}$ in the form 
$V^{++}\to V^{++}+g\,v^{++}$ \cite{bbko,bko}
with $V^{++}$ in the right-hand side being a  background  and
$v^{++}$ being a quantum superfield.
The  hypermultiplet will have no background, 
i.e. will be treated as  a quantum superfield.


As  explained  above, 
we  are interested  in the large $N$ part of  the 2-loop 
effective  action in  the case of the $U(1)$  background
\rf{backg},\rf{beck} corresponding 
to $N$ coincident 
 D3-branes separated by a distance $X$ 
 from  one D3-brane carrying $F_{mn}$
 field. The background matrix $J$ is the traceless part of the 
 $u(N+1)$ matrix  $ {\rm diag} (1, 0, ...,0) $.\foot{
Writtent in adjoint representation it  is just a combination 
of differences of its diagonal elements,  
$  {\rm diag} (0,... 0, 1,-1,...,1,-1) $, 
i.e. it contains $N^2+1$ zeros and $N$ pairs $(1,-1)$.}
Since we are after   the planar contribution  only 
we may just consider   the $U(N+1)$ theory   and ignore the subtraction of
traces in the background field 
expressions and  propagators.
The relevant planar 2-loop graphs (represented 
in double-line notation) will have the following structure
(see Fig.1):
 the background fields  will be 
attached only to the ``outer'' cycle  of the diagram
(representing, in string theory language, 
 the loop lying  on the single separated D3-brane), 
with two internal 
cycles (lying on $N$ coincident D3-branes)    each  producing a 
trace factor of $N$.\foot{In more general case when the probe is 
a cluster of $n$ D3-branes  the planar contribution
 will be proportional  to $N^2 \tr $(products of background matrices in
 fundamental representation).}

\hspace{5.5cm}
\Lengthunit=1.8cm
\GRAPH(hsize=3){
\mov(.5,0){\Circle(2)\Circle(1.8)
\mov(-1,0){
}
\mov(-1,.05){\lin(1.7,0)}\mov(-1,-.05){\lin(1.7,0)}
\mov(-.9,-.7){\lin(-.7,-.7)}\mov(-.8,-.7){\lin(-.7,-.7)}
\mov(.4,.7){\lin(.7,.7)}\mov(.5,.7){\lin(.7,.7)}
\mov(-.9,.7){\lin(-.7,.7)}\mov(-1,.7){\lin(-.7,.7)}
\mov(.3,-.7){\lin(.7,-.7)}\mov(.2,-.7){\lin(.7,-.7)}
\mov(-.7,1){\lin(0,1)}\mov(-.8,1){\lin(0,1)}
\mov(-.2,1){\lin(0,1)}\mov(-.1,.9){\lin(0,1.1)}
\ind(-18,6){ }\ind(-17,-6){ }\ind(-9,-19){Fig. 1}
}}

\bigskip





To develop  perturbation theory  one needs 
${\cal N}=2$ background dependent  superfield   propagators.
 They can be found by analogy with 
\cite{bko} 
\begin{eqnarray}
<v^{++}_\t(1)\,v^{++}_\t(2)>
&=& - \frac{2{\rm i}}{{\stackrel{\frown}{\Box}_1}{}}
{\stackrel{\longrightarrow}{(\cD_1^+)^4}}{}
\biggl\{ \delta^{12}(z_1-z_2)\delta^{(-2,2)}(u_1,u_2)  \biggr\}\ , 
\non \\
<q^{+i}_\t(1)\,\breve{q}^+_{j\t}(2)>
&=& \; \delta^i_j\frac{{\rm i}}{{\stackrel{\frown}{\Box}}{}}
{\stackrel{\longrightarrow}{({\cD}_1^+)^4}}{}
\left\{ \delta^{12}(z_1-z_2)
{1\over (u^+_1 u^+_2)^3} \right\}
{\stackrel{\longleftarrow}{({\cD}_2^+)^4}}\ , 
\non \\
<{ c}_\t(1)\,\,{ b}_\t(2)>
&=& - \frac{2{\rm i}}{{\stackrel{\frown}{\Box}}{}}
{\stackrel{\longrightarrow}{({\cD}_1^+)^4}}{}
\left\{ \delta^{12}(z_1-z_2)
{(u^-_1 u^-_2)\over (u^+_1 u^+_2)^3}
\right\}
{\stackrel{\longleftarrow}{({\cD}_2^+)^4}}\;.
\label{28}
\end{eqnarray}
The first line here defines the ${\cal N}=2$ SYM
propagator, the second -- hypermultiplet propagator and third 
-- the  ghost propagator. The index $\t$ means that the corresponding
superfields are taken in the  so called $\t$-frame \cite{GIKOS}. 

We have  suppressed the indices  of the fundamental representation, 
$v= (v)_{kl}$ ($k,l=1,..., N+1$), i.e.  the propagators  carry  
the two  pairs
of indices 
$kl, k'l'$.  In the absence of the background  each propagator 
is  proportional to 
$\delta_{kk'}\delta_{ll'}$ (dropping trace terms subleading  at large $N$).
In the presence of  our  $U(1)$  background  the 
propagator $P_{kk', ll'}$ 
will  remain to be diagonal,   with  
$P_{11,11} = P^{(0)}$ + background-dependent terms, 
$P_{1s,1s'} = \delta_{ss'} P^{(0)} + $background-dependent terms,
$P_{st, s't'} =  P^{(0)}_{st,s't'}$, where 
$s,t=2, ..., N+1$ are indices of the fundamental 
representation of the unbroken $SU(N)$ group
and $P^{(0)}_{kl,k'l'} = P^{(0)} \delta_{kl} \delta_{k'l'}$
is the free propagator.
Thus all one is to do is to separately  take into account 
 the two type of contractions  -- with  the $U(1)$ index  ``1''
 (which involve  the background-dependent propagator)
 and with  the $SU(N)$ indices $a,b$ taking valies $2,...,
 N$ (which involve  the
 free  propagator).  
The part $P_{11,11}$ will not contribute to the  leading large $N$ 
part of the diagram. It is  obvious  also that the only 
part of $P_{1a,1b'}$ that will be contributing to
the  relevant diagrams
(with all background dependence at outer line only)
will be 
$\bar P_{1a,1a'} = \delta_{aa'} P,$ \ 
 $P = P^{(0)} + $background-dependent terms.
It is effectively {\it this}   non-trivial background-dependent
block of the propagator that will  be discussed below.
The role of the remaining free $SU(N)$ index contractions  
is simply to produce  the   factor of $N^2$.


Since the  ${\cal N}=2$ background superfields ${\cW},\bar{\cW}$ will be   
on-shell \cite{bbk},  
i.e.  will be subject to  the equations of motion
(at the end, satisfying   \rf{backg})
\bea
\label{eqm}
{\cD}^{\a(i}\cD^{j)}_\a {\cW}=0 \ , 
\eea
and are also abelian,  one is able to 
show that $[{\cal D}^{+\alpha},{\cal D}^-_\alpha] {\cW}=0$, \ 
${\cal D}^{+\a} {\cal D}^+_\a {\cW}=0$.
Then  the operator ${\stackrel{\frown}{\Box}}$ \cite{bbko} in
(\ref{28})  takes the form
\begin{eqnarray}
{\stackrel{\frown}{\Box}}{}&=&
\Box+
\frac{{\rm i}}{2}({\cal D}^{+\alpha}{\cal W}){\cal
D}^-_\alpha+\frac{{\rm i}}{2} ({\bar{\cal D}}^+_{\dot\alpha}{\bar
{\cal W}}){\bar{\cal D}}^{-{\dot\alpha}}+
\bar {\cW}\cW\ . 
\label{24a}
\end{eqnarray}

%
Within the 
background field method,  the  2-loop 
corrections to the 
effective action  are given by the ``vacuum" diagrams containing only
cubic  and quartic vertices of quantum superfields. The  
supergraphs with {\it cubic}  vertices  have the form

\vspace{3mm}

\mov(2.5,0){\wavecirc(1.0)\mov(-0.5,0){\wavelin(1.0,0)}
\mov(2.0,0){\Circle(1.0)}\mov(2.5,0){\wavelin(-1.0,0)}
\mov(4.0,0){\dashcirc(1.0)}\mov(4.5,0){\wavelin(-1.0,0)}
\ind(5,-7){Fig.2a}\ind(27,-7){Fig.2b}\ind(47,-7){Fig.2c}
}

\vspace{2mm}
\noindent
Here the  wavy, solid and dashed lines stand for  the propagators
of  the ${\cN}=2$ gauge, hypermultiplet and ghost superfields
 respectively.
It is easy to see from (\ref{28}) that the  effective action can be
represented as a power series in supercovariant derivatives of the 
background superfields. 

To  find the leading   corrections to the 2-loop 
effective action
$\Gamma^{(2)}$ 
coming from these
supergraphs,  we  are to  represent  the
operator $\frac{1}{\stackrel{\frown}{\Box}}$ as a power series in
derivatives of the ${\cal N}=2$ background strength.
 This expansion looks like
\bea
\label{expa}
\frac{1}{\stackrel{\frown}{\Box}}=\frac{1}{\Box+{\cW}\bar{\cW}}
\sum_{n=0}^{\infty}\Big\{(-\frac{i}{2})
      [{\cal D}^{+\alpha}{\cal W}){\cal
D}^-_\alpha+({\bar{\cal D}}^+_{\dot\alpha}{\bar
{\cal W}}){\bar{\cal
D}}^{-{\dot\alpha}}]\frac{1}{\Box+{\cW}\bar{\cW}}\Big\}^n\ . 
\eea
Then we are to substitute this expansion 
into the expressions corresponding to the 
supergraphs in  Figs. 2a -- 2c and to carry out (covariant) 
$\cD$-algebra transformations (for a description of the
$\cD$-algebra   see 
\cite{BK0}). 
We will find  that to obtain  the leading contribution to the 
low-energy effective action 
one should keep  only the 
term which is  of fourth order in ${\cD}{\cW}$ and of second order in 
$\bar{\cD}\bar{\cW}$ (plus the conjugate term).
It  is this term  that,   when 
written in components,  contains the 
$F^6/|X|^8$ correction  we are interested in computing
(cf. \rf{twoo}).


Let us now consider the
 contribution of the  two-loop supergraph  in Fig. 2d containing 
 the {\it quartic}   vertex (which  is  presents only in  the ${\cal N}=2$
  gauge superfield sector).

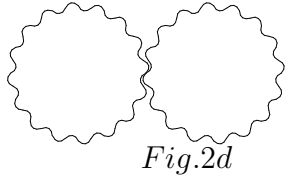
\vspace*{2mm}

\mov(3.0,0){\wavecirc(1)\mov(1.0,0){\wavecirc(1)}}
\ind(38,-1){Fig. 2d}

\vspace*{2mm}

It is possible to show 
that, in contrast to the  
supergraphs in Figs. 2a -- 2c,
 the   supergraph  in Fig. 2d 
 gives only   a sub-leading contribution 
to $\Gamma$ and thus may be ignored.
As usual,  the form of the superpropagators  implies 
that the  supergraphs contain
Grassmann delta-functions allowing  one to represent their
 contributions in the form 
 local in $\theta$-coordinates. Transformations leading to such form 
 may be  called ``contracting  loops into points in
Grassmann space''.
To contract   a single loop into a point we need 8 $\cD,\bar \cD$
factors. Any propagator
carries
4 manifest $\cD$-factors  and   additional $\cD,\bar \cD$ factors
may come  from the expansion of
$\stackrel{\frown}{\Box}^{-1}$. To contract two loops in Fig. 2d into
points
in $\theta$-space we thus need 16 $\cD,\bar \cD$
factors, with  8 of then coming from 
 the expansion of ${\stackrel{\frown}{\Box}}^{-1}$.
Any  $\cD$ or $\bar \cD$   factor from the  expansion of  
$\stackrel{\frown}{\Box}$ is accompanied  by 
$ {\cal D}{\cal W}$ or $  \bar {\cal D}\bar {\cal W} $
    factor.
Therefore,   a non-zero  contribution from 
the   supergraph in Fig. 2d will be   proportional to 
$ ({\cal D}{\cal W})^4   (\bar {\cal D}\bar {\cal W})^4$.
 Such   term is  subleading in the low-energy expansion of the 2-loop
 effective action  since, as we shall see, 
  the leading term  coming from the  cubic vertex  
 supergraphs in Figs. 2a -- 2c  is proportional to
$({\cal D}{\cal W})^4(\bar{\cal D}\bar{\cal W})^2+h.c.$
 

Let us now  discuss the structure 
of  propagators and vertices in our  background.
The quadratic part of the action 
of ${\cal N}=2$ gauge superfield  has  the form
\bea
\label{qu}
S_2=-\frac{1}{4}{\rm tr}\int d\xi^{(-4)} v^{++}\stackrel{\frown}{\Box} v^{++}
\eea
Since the  fields take values in the adjoint representation,
\bea
\label{ad}
\stackrel{\frown}{\Box} v^{++}=
\Box v^{++}+\frac{i}{2}[{\cal D}^{+\a}{\cal 
W},{\cal D}^-_{\a}v^{++}]+
\frac{i}{2}[\bar{\cal D}^{+\ad}{\bar{\cal
 W}},\bar{\cal D}^-_{\ad}v^{++}]+
[{\cal W}\bar{\cal W},v^{++}]
\eea
The  quantum vector field with  values in $u(N+1)$
can be written as 
$v^{++}=v^{++}_{kl}e_{kl}$, where  $e_{kl}$ 
is the Weyl basis of $u(N+1)$  ($k,l=1,...,N+1$)\footnote{
Since we are interested in the planar contribution 
we may not distinguish between $u(N+1)$ and $su(N+1)$.
Note also that 
since the superfield
 $v^{++}$ is real $v^{++}_{1a}=\bar{v}^{++}_{a1}$.}
\bea
\label{gen}
(e_{kl})_{pq}=\delta_{kp}\delta_{lq}\ .
\eea
The background strength  is then 
$
{\cal W}={\cal W}_{kl}e_{kl}.
$
In the case under consideration 
(where the background field takes values only in 
the unbroken 
 $u(1)$ which we label by index ``1'')
 we have: 
${\cal W}_{11}\equiv {\cal W}_0\neq 0$, with 
all other components ${\cal W}_{ij}$  equal to
zero. Let us denote the matrix  $e_{11}$ as $r$, i.e.
$r_{ab}=\delta_{1a}\delta_{1b}$, 
$a,b=2,...,N+1$.
Then the quadratic part of 
the action can be written as 
\bea
S_2
&=&-\frac{1}{4}
\int d\xi^{(-4)}\Big\{v^{++}_{kl}\Box v^{++}_{mn}\ {\rm tr}(e_{kl}e_{mn})+
\nonumber\\&+&
\frac{i}{2}v^{++}_{kl}\Big(
{\cal D}^{+\a} {\cW}_0 {\cal D}^-_{\a}
+ \bar{\cal D}^{+\ad}\bar{\cW}_0\bar{\cal D}^-_{\ad} \Big) 
v^{++}_{mn}
\ 
{\rm tr}(e_{kl}[r,e_{mn}])+\nonumber\\&+&
v^{++}_{kl}{\cW}_0\bar{\cW}_0 v^{++}_{mn}\ {\rm tr}[e_{kl}[r,[r,e_{mn}]]\Big\}
\ . \eea
Using  (\ref{gen}) we get 
\bea
\label{2fin}
S_2&=&-\frac{1}{4}\int d\xi^{(-4)}\Big\{
v^{++}_{11}\Box v^{++}_{11}+2v^{++}_{1a}\Box v^{++}_{a1}+
v^{++}_{ab}\Box v^{++}_{ba}+\nonumber\\
&+&2 (v^{++}_{a1}{\cal D}^{+\a} {\cW}_0
 {\cal D}^-_{\a}v^{++}_{1a}+h.c.)
+\nonumber\\&+&2v_{1a}v_{a1}{\cW}_0\bar{\cW}_0
\Big\}
\eea
From here we  conclude that only the 
components
$v_{1a},v_{a1}$ ($a\neq 1$) have a 
non-trivial background-dependent 
 propagator: all  other quantum field 
  components $va_{11}$ and $v_{ab}$ ($a,b\neq 1$)
do not interact with the background, i.e.  have the  free 
  propagator $\Box^{-1}$.

Repeating the same procedure for the 
 ghost and hypermultiplet fields we get the following set of 
 propagators 
\bea\label{koz}
<v^{++}_{1a}(z_1,u_1)\bar{v}^{++}_{1b}(z_2,u_2)>&=&-2i\delta_{ab}
\frac{({\cal D}^+_1)^4}{\stackrel{\frown}{\Box}_w}\delta^{12}_{12}
\delta^{(2,-2)}(u_1,u_2)\nonumber\\
<b_{a1}(z_1,u_1)c_{1b}(z_2,u_2)>&=&-2i\delta_{ab}
\frac{(u^-_1u^-_2)}{(u^+_1u^+_2)^3}
\frac{({\cal D}^+_1)^4({\cal D}^+_2)^4}
{\stackrel{\frown}{\Box}_w}\delta^{12}_{12}\nonumber\\
<\stackrel{\smile}{q}^+_{1a}(z_1,u_1)q^+_{b1}(z_2,u_2)>&=&2i\delta_{ab}
\frac{1}{(u^+_1u^+_2)^3}
\frac{({\cal D}^+_1)^4({\cal D}^+_2)^4}
{\stackrel{\frown}{\Box}_w}\delta^{12}_{12}\nonumber\\
<b_{1a}(z_1,u_1)c_{b1}(z_2,u_2)>&=&\delta_{ab}
\frac{(u^-_1u^-_2)}{(u^+_1u^+_2)^3}
\frac{({\cal D}^+_1)^4 ({\cal D}^+_2)^4}
{\stackrel{\frown}{\Box}_w}\delta^{12}_{12}\nonumber\\
<\stackrel{\smile}{q}^+_{a1}(z_1,u_1)q^+_{1b}(z_2,u_2)>&=&2i\delta_{ab}
\frac{1}{(u^+_1u^+_2)^3}
\frac{({\cal D}^+_1)^4 ({\cal D}^+_2)^4}
{\stackrel{\frown}{\Box}_w}\delta^{12}_{12}\ , 
\eea
where 
\bea
\stackrel{\frown}{\Box}_w&=&\Box+\frac{i}{2}[({\cal D}^{+\a}{\cW}_0)
{\cal D}^-_{\a}+(\bar{\cal D}^{+\ad}\bar{\cW}_0)\bar{\cal D}^-_{\ad}]+
{\cW}_0\bar{\cW}_0\ . 
\eea
Here 
the indices $a,b$ are { not} equal to 1.
All other components of the 
propagators are the same as in free theory
(cf. \cite{GIKOS}). 
Note that the propagators \rf{koz}
have 
extra factor 2 as  compared to the ones in \cite{GIKOS}
bacause of the extra factor
1/2 in the action \rf{sq}.

The leading $N\to\infty$
contribution from the supergraphs in  Figs. 1a -- 1c  
contains the following group-theory 
factor (i.e. the factor that multiplies the direct 
product of ``singlet'' propagators)\foot{The numerical 
 factor $ -{ 1 \ov 3} $ appears 
as follows:
each vertex carries the factor 
$\frac{1}{6}$, contracting
quantum fields into propagators gives  the factor 6,
two propagators $<v^{++}_{1a}\bar{v}^{++}_{1b}>$ contribute 
factor 4, the propagator 
$<v^{++}_{ab}v^{++}_{cd}>$ gives factor of 1, 
and the expansion of $\exp(iS_{int})$ in  the path integral
to the second order in  gauge coupling 
contributes the factor $-\frac{1}{2}$.}
$-\frac{1}{3}
{\rm tr}(e_{1a}[e_{d1},e_{bc}]){\rm tr}(e_{1a}[e_{d1},e_{bc}]).$ 
Using  (\ref{gen}) one finds that this 
factor is equal to 
$-\frac{1}{3}N^2$.

Below  we  shall omit the  index $w$  
on the background-dependent  operator   $ \stackrel{\frown}{\Box}_w
$; we will also  rename 
${\cW}_0$ and $\bar{\cW}_0$ as simply 
${\cal W}$ and $\bar{\cal
  W}$, which will stand for 
  the  non-zero
components  of ${\cal N}=2$ superfield strength.

\section{Hypermultiplet and ghost contributions\\
 to  2-loop 
low-energy effective action }
\setcounter{equation}{0}

Let us consider in detail the structure  of   supergraphs
corresponding to  Figs. 2a -- 2c.
We  may  first contract the 
 gauge, hypermultiplet and ghost loops  to a 
point in $\theta$-space using the rule \cite{bbko}\foot{We 
use the notation 
$ \delta^{12}_{12}= \delta^8_{12}\delta^4_{12}$ 
 for the $\N=2$  superspace
$\delta$-function of  argument $z_1-z_2$, with 
$ \delta^8_{12}$ being its Grassmann part.}
\bea
\label{point}
\delta^8_{12}{(\cD^+(u_1))}^4{(\cD^+(u_2))}^4\delta^8_{12}=
(u^+_1u^+_2)^4\delta^8_{12}\ . 
\eea
Then the leading  $N\to\infty$ 
contributions of these supergraphs to  
the effective action 
can be represented as
\bea
\label{co}
I_a&=&-\frac{1}{3}g^2_{\rm YM} N^2 
\int d^8\theta_1 d^8\theta_2 du_1  du_2 dv_2 dw_2
\frac{{(D^+)}^4(u_1)}{\Box}
\delta^{12}_{12}
\delta^{(2,-2)}(u_1,u_2)
\nonumber\\&\times&
\frac{(v^+_2w^+_2)^2}
{(u^+_1v^+_2)(u^+_1w^+_2)(u^+_2v^+_2)(u^+_2w^+_2)}
\frac{1}{\stackrel{\frown}{\Box}(v_2)}\delta^{12}_{12}
\frac{1}{\stackrel{\frown}{\Box}(w_2)}
\delta^4(x_1-x_2)\nonumber\\
I_b&=&-\frac{1}{3}g^2_{\rm YM} N^2\int d^8\theta_1 d^8\theta_2 du_1 du_2
\frac{{(D^+)}^4(u_1)}{\Box}
\delta^{12}_{12}
\delta^{(2,-2)}(u_1,u_2)\non\\&\times&
\frac{1}
{(u^+_1u^+_2)^2}
\frac{1}{\stackrel{\frown}{\Box}(u_1)}\delta^{12}_{12}
\frac{1}{\stackrel{\frown}{\Box}(u_1)}\delta^4(x_1-x_2)
\nonumber\\
I_c&=&-\frac{2}{3}g^2_{\rm YM} N^2\int d^8\theta_1 d^8\theta_2 du_1 du_2
\frac{{(D^+)}^4(u_1)}{\Box}
\delta^{12}_{12}
\delta^{(2,-2)}(u_1,u_2)\nonumber\\&\times&
D^{++}_1D^{++}_2
\big\{\frac{(u^-_1u^-_2)^2}{(u^+_1u^+_2)^2}
\frac{1}{\stackrel{\frown}{\Box}(u_1)}\delta^{12}_{12}
\frac{1}{\stackrel{\frown}{\Box}(u_1)}
\delta^4(x_1-x_2)\big\}
\eea
Here $D^+$ is the  ``flat'' derivative originating 
from the free propagator. 
To obtain these expressions we note that product of
the  three propagators
(\ref{28}) carries factor  $i$, and 
the two-loop correction $\Gamma^{(2)}$
is defined with  factor $i$ ($Z= e^{i\Gamma}$). 
It was already mentioned that due to
the  structure of the vertices one
propagator in each supergraph is the free one.

In  $I_a$ we may  write $(v^+_2w^+_2)^2=D^{++}_{v_2}D^{++}_{w_2}
\{(v^+_2w^+_2)(v^-_2w^-_2)\}$, and integrating by parts, 
transport the derivatives $D^{++}_{v_2},
D^{++}_{w_2}$ to  the  other part of integrand. We 
may also do the same operation
with harmonic derivatives in  
$I_c$. 
These transformations are completely analogous to the ones
 done in  \cite{high} in the case of constant
  ${\cW},\bar{\cW}$ 
 (where all the  derivatives of 
 the background fields ${\cW},\bar{\cW}$ were zero).  
However, here  (cf. \rf{backg}),
  unlike  the  case  of the 
   nonholomorphic effective potential 
  considered     in \cite{high}, 
the factor  $\frac{1}{\stackrel{\frown}{\Box}}$ 
depends on the  harmonic coordinates  (since 
 ${\cW},\bar{\cW}$  are  not constant in superspace 
 this dependence is implied by \rf{expa}).
 Therefore,
$I_{a,c}$ contain additional terms where $D^{++}$'s act on 
$\frac{1}{\stackrel{\frown}{\Box}}$, we get 
\bea
I_a&=&\frac{2}{3}
g^2_{\rm YM} N^2\int d^8\theta_1 d^8\theta_2 du_1  du_2 dv_2 dw_2
\frac{{(\cD^+)}^4(u_1)}{\Box}
\delta^{12}_{12}
\delta^{(2,-2)}(u_1,u_2) (v^+_2w^+_2)(v^-_2w^-_2)             
\nonumber\\&\times&
D^{++}_{v_2}D^{++}_{w_2}\Big[\frac{1}
{(u^+_1v^+_2)(u^+_1w^+_2)(u^+_2v^+_2)(u^+_2w^+_2)}
\frac{1}{\stackrel{\frown}{\Box}(v_2)}\delta^{12}_{12}
\frac{1}{\stackrel{\frown}{\Box}(w_2)}\delta^4(x_1-x_2)\Big]\nonumber\\
I_c&=&\frac{4}{3}g^2_{\rm YM} N^2\int d^8\theta_1 d^8\theta_2 du_1 du_2
\frac{{(\cD^+)}^4(u_1)}{\Box}
\delta^{12}_{12}
\delta^{(2,-2)}(u_1,u_2)\nonumber\\&\times&
D^{++}_1D^{++}_2
\Big[\frac{(u^-_1u^-_2)^2}{(u^+_1u^+_2)^2}
\frac{1}{\stackrel{\frown}{\Box}(u_1)}\delta^{12}_{12}
\frac{1}{\stackrel{\frown}{\Box}(u_1)}\delta^4(x_1-x_2)\Big]
\eea
The contributions of  each of these supergraphs $I_{a,b,c}$
can be separated into 
two parts, $I= I' + I''$. 
 The first corresponds to
the terms in which  $D^{++}$'s do not act on
$\frac{1}{\stackrel{\frown}{\Box}}$.
 The second one contains 
the terms where  $D^{++}$'s are acting on $\stackrel{\frown}{\Box}$.
Note that $I_b$ contains only 
 the contribution of the  first type.
The sum of contributions in which $D^{++}$'s  do not
act on $\stackrel{\frown}{\Box}$ vanishes identically due to
$\N=4$ supersymmetry,  as it was pointed out in \cite{high}.
That  means that {\it  diagrams  with
 hypermultiplet  propagators do not contribute to the 
leading  2-loop term in the low-energy effective
action.}

What remains is to  consider  the terms in $I_{a,c}$ which
contain   $D^{++}$'s  acting on $\stackrel{\frown}{\Box}$. 
Such  term  in $I_c$ is 
\bea
\label{sumsp12}
I^{''}_c&=&-\frac{4}{3}g^2_{\rm YM} N^2 \int d^8\theta\int
du_1du_2
\frac{{(\cD^+)}^4(u_1)}{\Box}
\delta^{12}_{12}
\delta^{(2,-2)}(u_1,u_2)
\Big[
\frac{(u^-_1u^-_2)(u^-_1u^+_2)}{(u^+_1u^+_2)^2}
\frac{1}{\stackrel{\frown}{\Box}(u_1)}\delta^{12}_{12}
\nonumber\\&\times&
\frac{1}{\stackrel{\frown}{\Box}(u_1)}
(({\cal D}^{+\alpha}(u_1){\cal W}){\cal D}^+_\alpha(u_1)+
({\bar{\cal D}}^+_{\dot\alpha}(u_1){\bar
{\cal W}}){\bar{\cal D}}^{+{\dot\alpha}}(u_1))
\frac{1}{\stackrel{\frown}{\Box}(u_1)}\delta^4(x_1-x_2)\Big]
\eea
The supercovariant derivatives of ${\cN}=2$ 
field strengths arise from 
the expansion of $\stackrel{\frown}{\Box}^{-1}$. 
In the above expression for  $I^{''}_c$ 
all factors $\stackrel{\frown}{\Box}^{-1}$ depend on the same harmonic 
coordinate $u_1$. Hence all derivatives ${\cD}^+{\cW}$ which appeared from
$\stackrel{\frown}{\Box}^{-1}$ also depend only
on $u_1$. Thus, all contributions generated by $I_c$ should be proportional
to various powers of 
${\cD}^+(u_1){\cW}$, $\bar{\cD}^+(u_1)\bar{\cW}$.
However, as one can show, 
all contributions of second order in
${\cD}^+(u_1){\cW}$, $\bar{\cD}^+(u_1)\bar{\cW}$ are equal to zero.
Also,  
$({\cD}_{\a}^+(u_1){\cW})^n=0$  for  $n\geq 3$.
Hence we 
 find that {\it diagrams with  ghost propagators 
  also do not contribute
  to the leading term in the 2-loop 
  low-energy effective action. }
This result 
 is similar  to the one known    in the 
   the one-loop
approximation where the   ghosts and  the 
  hypermultiplets  also 
do not contribute to the on-shell
 low-energy effective action \cite{bbk}.

The term in  $I_a$ containing  $D^{++}$  acting on 
$\stackrel{\frown}{\Box}$ takes the form
(after  integrating  over $u_2$ and some transformations) 
\bea
\label{sumsp11b}
I^{''}_a&=&-\frac{2}{3}g^2_{\rm YM} N^2\int d^8\theta\int
dudvdw
\frac{{({\cD}^+)}^4(u)}{\Box}
\delta^{12}_{12}
\nonumber\\& & \times
\Big[
-\frac{1}{4}
\frac{(v^-w^-)(v^+w^+)}{(u^+v^+)^2(u^+w^+)^2}
\frac{1}{\stackrel{\frown}{\Box}(v)}(
({\cal D}^{+\alpha}(v){\cal W}){\cal
D}^+_\alpha(v)+({\bar{\cal D}}^+_{\dot\alpha}(v){\bar
{\cal W}}){\bar{\cal D}}^{+{\dot\alpha}}(v)
)\frac{1}{\stackrel{\frown}{\Box}(v)}\delta^{12}_{12}
\nonumber\\& & \times
\frac{1}{\stackrel{\frown}{\Box}(w)}(
({\cal D}^{+\alpha}(w){\cal W}){\cal
D}^+_\alpha(w)+({\bar{\cal D}}^+_{\dot\alpha}(w){\bar
{\cal W}}){\bar{\cal D}}^{+{\dot\alpha}}(w)
)\frac{1}{\stackrel{\frown}{\Box}(w)}\delta^4(x_1-x_2)\nonumber\\
& &
-\ i[D^{--}_v\delta^{(1,-1)}(u,v)]\frac{1}{\stackrel{\frown}{\Box}(v)}
\delta^{12}_{12}
\frac{1}{(u^+w^+)^2}
\frac{1}{\stackrel{\frown}{\Box}(w)}
(({\cal D}^{+\alpha}(w){\cal W}){\cal
D}^+_\alpha(w)\nonumber\\& & + \ 
({\bar{\cal D}}^+_{\dot\alpha}(w){\bar
{\cal W}}){\bar{\cal D}}^{+{\dot\alpha}}(w))
\frac{1}{\stackrel{\frown}{\Box}(w)}\delta^4(x_1-x_2)\Big]
\eea
The remaining  task   is  to  extract  from here  
the leading low-energy 
contribution to the 2-loop effective action.

\section{
Leading low-energy  term in \\
  ${\cal N}=2$  gauge field 2-loop contribution
 }

\setcounter{equation}{0}

To develop  low-energy  expansion of the  gauge field contribution
 $I^{''}_a$ \rf{sumsp11b} we use  (\ref{expa})
to represent  the factors
$\frac{1}{\stackrel{\frown}{\Box}}$ in (\ref{sumsp11b}) 
as  power series in
derivatives of ${\cW},\bar{\cW}$.
As a result, eq. (\ref{sumsp11b}) will be given by a 
 series of terms containing some number of covariant derivatives acting on
${\cW},\bar{\cW}$ and some powers of $(\Box+{\cW}\bar{\cW})^{-1}$.
The first term in this series has four derivatives and
 four ${\cal W}$  factors -- 
it is  this term that  determines the 
 leading  correction  to the two-loop
effective action.
It contains also the factor of 
$(\bar{\cD}\bar{\cW})^2$, so that  the leading contribution 
is proportional to 
\be
({\cD}{\cW})^4(\bar{\cD}\bar{\cW})^2 + h.c. \ , \la{oooi}
\ee
 and, 
when written
in components,  
 contains the 
 required  $F^6$ term in  the bosonic sector.
 
The contraction of the  remaining loop  to a  point
 using  the rule (\ref{point}) requires
four $\cD$-factors and four $\bar{\cD}$-factors. Therefore,
 we must take into
account only those terms in the  expansion which are at
 least of fourth order in
${\cD}{\cW}{\cD}$ and of second order in
$\bar{\cD}\bar{\cW}\bar{\cD}$. 
The only non-vanishing  supergraph producing 
 such  leading correction is given in  Fig.
3.
All other supergraphs are 
proportional to either  $(u^-u^-)=0$ or 
 $({\cD}^+(u){\cW})^3=0$
(this can be shown using 
 the
transformations similar to ones which are carried out below).
We shall omit also all commutators of the spinor supercovariant
 derivatives 
with $\Box+{\cW}\bar{\cW}$ since they  lead only to
 subleading 
corrections.


\hspace{5.5cm}
\Lengthunit=1.8cm
\GRAPH(hsize=3){
\mov(.5,0){\wavecirc(2)\mov(-1,0)
{
\wavelin(2,0)
}
\mov(-.8,-.7){\wavelin(-.7,-.7)}
\mov(.6,.7){\wavelin(.7,.7)}\mov(-.9,.7){\wavelin(-.7,.7)}
\mov(.5,-.7){\wavelin(.7,-.7)}
\mov(-.5,1){\wavelin(0,1)}\mov(.2,.8){\wavelin(0,1)}
\ind(10,-17){\cD^{+\beta}\cW} \ind(-8,22){\bar{\cD}^{+\dot{\beta}}\bar{\cW}}
\ind(6,-6){\cD^+_{\beta}} \ind(-20,-6){{\cD}^-_{\delta}}
\ind(-25,-17){\cD^{+\delta}\cW}\ind(-25,17){\cD^{+\alpha}\cW}
\ind(0,20){\bar{\cD}^{+\dot{\alpha}}\bar{\cW}}\ind(10,17){\cD^{+\gamma}\cW}
\ind(-8,9){|}\ind(-13,9){|}
\ind(-15,-1){|}\ind(-6,-1){|}\ind(-15,2){{(\cD^+)}^4}\ind(-6,2){{(\cD^+)}^4}
\ind(-2,6){-}\ind(-1,-6){-}
\ind(-18,6){-}\ind(-17,-6){-}\ind(-9,-19){Fig. 3}
\ind(2,3){{(\cD^+)}^4}\ind(-2,3){-}
\ind(0,8){\cD^-_{\gamma}}\ind(-11,13){\bar{\cD}^-_{\dot{\alpha}}}
\ind(-24,8){\cD^-_{\alpha}}\ind(-19,13){\bar{\cD}^-_{\dot{\beta}}}
}}

\bigskip 

The  contribution of the supergraph in Fig. 3 (after contraction of one
of the loops  to a point in $\q$-space as  in section 3)
 is given by 
\bea
\label{f60}
\Gamma^{(2)}&=&- 4g^2_{\rm YM} N^2
\int d^{12}z_1 d^{12}z_2 \int du dv dw\ 
 A(u,v,w)
\frac{1}{(\Box+{\cW}\bar{\cW})^5}\delta^{12}_{12}
({\cal D}^{+\a}_1(v){\cal W}){\cal D}^-_{1\a}(v)
\nonumber\\&\times&
({\cal D}^{+\g}_{1}(v){\cal W}){\cal D}^-_{1\g}(v)
({\cal D}^{+\b}_1(w){\cal W}){\cal D}^+_{1\b}(w)
({\cal D}^{+\delta}_{1}(w){\cal W}){\cal D}^+_{1\delta}(w)
\nonumber\\&\times&
(\bar{\cal D}^+_{1\ad}(w)\bar{\cal W})\bar{\cal D}^-_{\ad 1}(w)
(\bar{\cal D}^+_{1\bd}(w)\bar{\cal W})\bar{\cal D}^{-\bd}_1(w)
(D^+_2(u))^2
(\bar{D}^+_2(u))^2\frac{1}{\Box}
\delta^{12}_{12}\nonumber\\&\times&
\frac{1}{(\Box+{\cW}\bar{\cW})^3}\delta^4(x_1-x_2)+h.c.
\eea  
Here $A(u,v,w)$ is a function of harmonics 
\bea
\label{harm}
A(u,v,w)=-\frac{1}{4}\frac{(v^-w^-)(v^+w^+)}{(u^+v^+)^2(u^+w^+)^2}\ . 
\eea
As above, ${\cal D}$ is the background dependent 
 derivative, and $D$ is the flat 
one.
We took into account that there are 6 supergraphs of the form similar
to that one given in Fig. 2 with difference 
being only 
in position of
external superfield strength insertions. Each of these supergraphs gives the
same contribution.
The derivatives in 
$({\cal D}{\cal W})$ and $(\bar{\cal D}\bar{\cal W})$ act only on 
${\cal W},\bar{\cal W}$. All other derivatives act on
all terms to  the right.
Since we  are interested in the background \rf{backg} we may 
 use  the  property  $\pa_m{\cal W}=\pa_m\bar{\cal W}=0$.
The equations of motion (\ref{eqm})
lead to ${({\cal D}^{\pm})}^2(v){\cal W}=0$ (here $v$ is an arbitrary
harmonic coordinate). Using this identity,
 we conclude 
that  for  on-shell $\pa_m$-constant background field 
 strengths no more than
one spinor derivative  can act on the background field. 
As a result,  we can transport
all ``excess'' spinor derivatives (additional to 
 the ones incorporated
in 
$({\cal D}{\cal W})$ and $(\bar{\cal D}\bar{\cal W})$)
 to  act on  the same Grassmann
$\delta$-function. 
This  allows us to rewrite (\ref{f60}) in the form
\bea
\label{f6}
\Gamma^{(2)}&=&- 4g^2_{\rm YM} N^2  \int d^{12}z_1 d^{12}z_2 \int du dv dw
A(u,v,w)\frac{1}{(\Box+{\cW}\bar{\cW})^5}\delta^{12}_{12}
\nonumber\\&\times&
({\cal D}^{+\a}_1(v){\cal W})({\cal D}^+_{1\a}(v){\cal W})
({\cal D}^{+\a}_1(b){\cal W})({\cal D}^+_{1\b}(v){\cal W})
(\bar{\cal D}^+_{1\ad}(w)\bar{\cal W})(\bar{\cal D}^{+\ad}_1(w)\bar{\cal W})
\nonumber\\&\times&
{\cal D}^{+\g}_1(v){\cal D}^-_{\g 1}(v)
{\cal D}^{+\delta}_1(w){\cal D}^-_{\delta 1}(w)
\bar{\cal D}^-_{\bd 1}(w)\bar{\cal D}^{-\bd}_1(w)(D^+_2(u))^2
(\bar{D}^+_2(u))^2\frac{1}{\Box}
\delta^{12}_{12}\nonumber\\&\times&
\frac{1}{(\Box+{\cW}\bar{\cW})^3}\delta^4(x_1-x_2)+h.c.
\eea
Now we can integrate over $\q_2$ using the rule
$\int d^4 x_2 d^8\theta_2 f(\theta_2,\ldots)=\int d^4 x_2 
D^4_2\bar{D}^4_2 
f(\theta_2,\ldots)|_{\theta_2=0}$. Here dots  stand 
for all arguments except 
$\theta_2$, and $D^4_2=D_2^{i\a}D^j_{2\a}
D_{2i}^{\b}D_{2j\b}$,  where $i,j$ are the indices
numerating  ${\cal N}=2$ spinor derivatives.
The equations of motion (\ref{eqm}) lead to the 
following on-shell identity
$
\bar{D}^+_{\bd 1}(w)\bar{\cal W}\bar{D}^{+\bd}_1(w)
\bar{\cal W}=\frac{1}{2}\bar{\cal D}^{+2}(w)\bar{\cal W}^2$.
Using this identity we rewrite
(\ref{f6}) as
\bea
\label{f6a}
\Gamma^{(2)}&=&- 2g^2_{\rm YM} N^2
\int d^4x_1 d^4x_2 d^8\theta_1 \int du dv dw A(u,v,w)
({\cal D}^+_1(v){\cal W})^2({\cal D}^+_1(w){\cal W})^2
\nonumber\\&\times&
(\bar{\cal D}^+_1(w))^2\bar{\cal W}^2
{\cal D}^4_2\bar{\cal D}^4_2\Big[\frac{1}{(\Box+{\cW}\bar{\cW})^5}
\delta^{12}_{12}
{\cal D}^{+\g}_1(v){\cal D}^-_{\g 1}(v)
{\cal D}^{+\delta}_1(w){\cal D}^-_{\delta 1}(w)
\bar{\cal D}^-_{\bd 1}(w)\bar{\cal D}^{-\bd}_1(w)
\nonumber\\&\times&
({\cal D}^+_2(u))^2(\bar{\cal D}^+_2(u))^2\frac{1}{\Box}
\delta^{12}_{12}
\Big]|_{\theta_2=0}
\frac{1}{(\Box+{\cW}\bar{\cW})^3}\delta^4(x_1-x_2)+h.c.
\eea
The integrand in (\ref{f6a}) is  local in Grassmann coordinates, 
 as  usual in  superfield theories (see, e.g.,  \cite{BK0}).

The integrand here  is evaluated at $\q_2=0$, i.e. it  
depends only on  $\q_1$.
To simplify (\ref{f6a}) it is convenient to transport $\bar{\cal
D}^2$ from $\bar{\cal W}^2$ to the rest of the factors -- this 
 allows us to get an  expression in which the 
numbers of chiral and anti-chiral derivatives acting on delta 
functions are
equal to each other. Since
the  background  is abelian, 
the gauge covariant derivatives acting on the 
background strengths are equivalent to the ``flat''  ones, 
${\cal D}^+_{\a}(v){\cal W}=D^+_{\a}(v){\cal W}$.
We integrate by parts using the rule
$\int d^8\q_1 (\bar{D}^2_1\bar{\cal W}^2) Y =\int d^8\q_1 
\bar{\cal W}^2 \bar{D}^2_1 Y $.
The action of $\bar{D}$ on ${\cal D}{\cal W}=  D {\cal W}$ 
leads to the space-time derivatives of the 
background field strength, i.e. to the terms 
 which  we  are  ignoring here.
As a result,  we arrive at
\bea
\label{f6b0}
\Gamma^{(2)}&=&-2g^2_{\rm YM} N^2   \int d^4x_1 d^4x_2 d^8\theta_1
\int du dv dw A(u,v,w)
({\cal D}^+_1(v){\cal W})^2({\cal D}^+_1(w){\cal W})^2
\bar{\cal W}^2\nonumber\\&\times&
\bar{D}^{+2}_1(w)
{\cal D}^4_2\bar{\cal D}^4_2\Big[\frac{1}{(\Box+{\cW}\bar{\cW})^5}
\delta^{12}_{12}
{\cal D}^{+\g}_1(v){\cal D}^-_{\g 1}(v)
{\cal D}^{+\delta}_1(w){\cal D}^-_{\delta 1}(w)
\bar{\cal D}^-_{\bd 1}(w)\bar{\cal D}^{-\bd}_1(w)
\nonumber\\&\times&
({\cal D}^+_2(u))^2
(\bar{\cal D}^+_2(u))^2
\frac{1}{\Box}
\delta^{12}_{12}
\Big]|_{\theta_2=0}
\frac{1}{(\Box+{\cW}\bar{\cW})^5}\delta^4(x_1-x_2)+h.c.
\eea
Now  we are  to evaluate the result of applying   ten covariant 
derivatives to the  factor in the square brackets in (\ref{f6b0}). 
  Fortunately, only a few terms in the final 
  complicated expression  are actually 
  contributing  to  the  term \rf{oooi}
  in the low-energy effective action we are interested in.
   To extract them 
let us consider the component  form 
of this expression.
The component content of the  term $({\cal D}_1{\cal W})^4\bar{\cal W}^2$ 
in pure gauge sector is
$F^6\q^4_1\bar{\q}^4_1+\ldots$. 
Therefore,  all dependence of the integrand in (\ref{f6b0})
 on $\q_1$ is
concentrated in
$({\cal D}^+_1(v){\cal W})^2({\cal D}^+_1(w){\cal W})^2
\bar{\cal W}^2, 
$
and to  find  the contribution of  (\ref{f6b0}) it is enough
to study the component 
\bea
\label{prod}
R&=&\bar{D}^{+2}_1(w){\cal D}^4_2\bar{\cal D}^4_2\Big[
\delta^{12}_{12}
{\cal D}^{+\g}_1(v){\cal D}^-_{\g 1}(v)
{\cal D}^{+\delta}_1(w){\cal D}^-_{\delta 1}(w)
\bar{\cal D}^-_{\bd 1}(w)  \nonumber\\&\times&
\bar{\cal D}^{-\bd}_1(w)({\cal D}^+_2(u))^2
(\bar{\cal D}^+_2(u))^2
\frac{1}{\Box}\delta^{12}_{12}
\Big]|_{\theta_1,\theta_2=0}\ , 
\eea
i.e. the term which is 
 of zeroth order in  both  $\q_1$ and $\q_2$. 
Then eq. (\ref{f6b0}) can be written as
\bea
\label{f6b}
\Gamma^{(2)}&=&-2g^2_{\rm YM} N^2 \int d^4x_1 d^4x_2 d^8\theta_1
 \int du dv dw A(u,v,w)
({\cal D}^+_1(v){\cal W})^2({\cal D}^+_1(w){\cal W})^2
\bar{\cal W}^2 \nonumber\\&\times& \frac{1}{(\Box+{\cW}\bar{\cW})^5} R
\frac{1}{(\Box+{\cW}\bar{\cW})^3}\delta^4(x_1-x_2)+h.c.\ . 
\eea
To obtain a non-trivial 
contribution  from  (\ref{prod}) we should act 
with at least four ${\cal D}$  and four $\bar{\cal D}$
derivatives on each of Grassmann 
$\delta$-function 
(otherwise we get terms of first and higher orders in
$\q,\bar{\q}$'s  which vanish at $\q_1=\q_2=0$).
Since all derivatives now act on $\delta$-functions which are  symmetric
with respect to the indices 1 and 2 we  may arrange all of them 
 acting on the first  argument $z_1$.  
We then find that the 
zeroth order in $\q_1$ in (\ref{prod}) corresponds to acting
by  at least
eight spinor derivatives on each of the  $\delta$-functions.
The only non-zero term in $R$  (\ref{prod}) is then 
\bea
\label{r}
\delta^4(x_1-x_2)
&\Big[&\bar{D}^{+2}(w)
{\cal D}^{+\g}(v){\cal D}^-_{\g}(v)
{\cal D}^{+\delta}(w){\cal D}^-_{\delta}(w)
\bar{\cal D}^-_{\bd}(w)\bar{\cal D}^{-\bd}(w)({\cal D}^+(u))^2
(\bar{\cal D}^+(u))^2\nonumber\\&\times&
\frac{1}{\Box}
\delta^{12}_{12}\  \Big]|_{\q_1,\q_2=0}.
\eea
Other terms are either of odd order in Grassmann
coordinates (and hence vanish at $\q_1=\q_2=0$) or proportional to 
$(w^+w^+)=0$.


Next, we are to  contract the remaining loop into a point in 
$\q$-space. To do  this we 
commute ${\cal D}$ and $\bar{\cal D}$ factors in (\ref{r})
using the identity
$\{{\cal D}^{\pm}_{\a}(v),\bar{\cal D}^{\pm}(w)_{\ad}\}=2i(v^{\pm}w^{\pm})
{\cal D}_{\a\ad}$.
As a result,  we obtain a sum of terms in which two ${\cal D}$-
and two $\bar{\cal D}$-factors form $\Box$. 
\foot{Schematically,  these transformations can be summarized  as follows.
We consider (\ref{r}), choose two derivatives ${\cal D}$ and 
two derivatives
$\bar{\cal D}$ and replace them by the factor 
$\Box$
multiplied by a product of harmonic arguments 
of chosen derivatives. Then we
add together  all 
 such terms corresponding to all possible choices of pairs
${\cal D}$ and $\bar{\cal D}$.} 
As a result,  we arrive at
the following expression for (\ref{r}):
\bea
R&=&
\delta^4(x_1-x_2)
[(u^+v^-)(u^+v^+)(u^+w^+)(u^+w^-)+(u^+v^-)(u^+v^+)(u^+w^+)^2(u^+w^-)(v^-w^-)
\nonumber\\&+&(u^+v^-)(u^+v^+)(u^+w^-)^2(v^+w^-)+
(u^+v^+)(u^+w^+)^3(w^-v^-)\nonumber\\&+&(u^+v^+)(u^+w^+)^2(u^+w^-)(v^-w^-)+
(u^+v^-)(u^+v^+)(u^+w^+)(u^+w^-)^2(v^+w^+)]
\nonumber\\&\times&
\Box
\frac{1}{\Box}\delta^4(x_1-x_2)\ . 
\eea
The factor $\Box$  thus  cancels against the 
same  factor  in the denominator. 

Substituting the above $R$ into (\ref{f6b}), doing Fourier
transform, and integrating  over $x_2$ 
one obtains $\Gamma^{(2)}$ (\ref{f6b}) in the form:
\bea
\label{f6b3}
\Gamma^{(2)}&=&-2g^2_{\rm YM} N^2   \int d^4x d^8
\theta\int du dv dw A(u,v,w)
\Big[
4(u^+v^+)(u^+w^+)(u^+w^-)^2(w^+v^-)\non\\&+&4
(u^+w^+)(u^+w^-)^3(w^+v^-)(w^+v^+)\non\\&+&4(u^+w^+)(u^+w^-)^2(u^+v^-)(w^+v^+)
-
4(u^+w^+)^2(u^+w^-)(u^+v^+)(w^+v^-)\non\\&+&4
(u^+w^+)^3(u^+w^-)(w^-v^-)(w^-v^+)-4(u^+w^+)^2(u^+w^-)(u^+v^-)(w^-v^+)^2
\nonumber\\&-&
4(u^+v^+)(u^+w^+)(u^+w^-)(u+v^-)+4
(u^+w^+)^2(u^+w^-)^2(w^+v^-)(w^-v^+)\non\\&+&
4(u^+w^+)^2(u^+w^-)^2(w^-v^-)(w^+v^+)
+
4(u^+v^+)(u^+w^+)^2(u^+w^-)(w^-v^-)\non\\&+&4
(u^+w^+)^2(u^+w^-)(u^+v^-)(w^-v^+)-4(u^+w^+)(u^+w^-)^2(u^+v^+)(w^+v^-)
\nonumber\\&-&
4(u^+v^-)(u^+w^+)(u^+w^-)^2(w^+v^+)+6
(u^+w^+)(u^+w^-)^2(u^+v^+)(w^+v^-)\non\\&+&6(u^+w^-)^2(u^+v^-)(u^+w^+)(w^+v^+)
+
3(u^+w^+)(u^+w^-)^3(w^+v^-)(w^+v^+)\non\\&+&3
(u^+w^+)(u^+w^-)(u^+v^-)(u^+v^+)+
3(u^+w^+)^2(u^+w^-)^2(w^-v^-)(w^+v^+)\non\\&+&
3(u^+w^+)^2(u^+w^-)^2(w^+v^-)(w^-v^+)
\Big]
\nonumber\\&\times&
({\cal D}^+(v){\cal W})^2({\cal D}^+(w){\cal W})^2
\bar{\cal W}^2 
\int\frac{d^4 k d^4 l}{(2\pi)^8}
\frac{1}{(k^2+{\cW}\bar{\cW})^5}\frac{1}{(l^2+{\cW}\bar{\cW})^3}
+h.c.
\eea
Here we  renamed the coordinates  as  $x_1=x,\, \q_1=\q$.
The momentum integrals can be easily calculated:
\bea
\label{ii}
\int\frac{d^4k d^4l}{(2\pi)^8}\frac{1}{(l^2+{\cW}\bar{\cW})^{3}
(k^2+{\cW}\bar{\cW})^{5}}=\frac{1}{(4\pi)^4}
\frac{1}{({\cal W}\bar{\cal W})^4}
\frac{\Gamma(1)}{\Gamma(3)}\frac{\Gamma(3)}{\Gamma(5)}=
\frac{1}{24(4\pi)^4{({\cal W}\bar{\cal W}})^4}\ . 
\eea
It remains to  integrate over the  harmonics. We use the  identities
like $(u^+w^+)=D^{++}_u(u^-w^+)$ and 
transport the harmonic derivatives $D^{++}$ to other factors
in the 
integrand. The  identity
$
D^{++}_u\frac{1}{(u^+v^+)}$=$\delta^{(0,0)}(u,v)
$
produces harmonic $\delta$-functions 
which  allow to do the  integral over $u$. As a result, after
substituting
of integral (\ref{ii}) into (\ref{f6b3}) we get
\bea
\label{theres}
\Gamma^{(2)}&=&  \frac{1}{48(4\pi)^4}g^2_{\rm YM} N^2\int d^{12}z\int dv dw 
({\cal D}^+(v){\cal W})^2({\cal D}^+(w){\cal W})^2 \bar{\cal W}^2
\frac{1}{({\cW}\bar{\cW})^4}\nonumber\\&\times&
(v^-w^-)^2[2+2(v^-w^-)(v^+w^+)-4(v^-w^-)(v^+w^+)(v^+w^-)(v^-w^+)
\nonumber\\&-&4(v^+w^-)(v^-w^+)-6(v^-w^+)^2(v^+w^-)^2-3
(v^-w^-)^2(v^+w^+)^2]
\eea
Calculating
the  integrals over the harmonics using the rules of 
\cite{GIKOS} (see Appendix B) gives 
(here 1, 2  are indices of the two  $\N=1$ Grassmann coordinates)
\bea
\label{res0}
\Gamma^{(2)}&= & \frac{1}{48(4\pi)^4}g^2_{\rm YM} N^2 \int d^{12}z\ 
\bar{\cal W}^2({\cal D}_1{\cal W})^2({\cal D}_2{\cal W})^2
\frac{1}{({\cW}\bar{\cW})^4} + h.c. \ .
\eea 
Using the equations of motion (\ref{eqm})
for ${\cal W},\bar{\cal W}$ we obtain the following final 
$\N=2$ supersymmetric expression 
for the leading part of the 2-loop $\cal N$=4  
SYM effective action  
\bea
\label{res}
\Gamma^{(2)}=  \frac{1}{48 (4\pi)^4}g^2_{\rm YM} N^2 
\int d^{12}z\  \frac{1}{\bar{\cW}^2}
\ln \frac{\cW}{\m}{\cD}^4
\ln \frac{\cW}{\m}+ h.c.\ . 
\eea
This  expression 
matches  the one in \rf{twoo} --  it reproduces 
exactly 
the  value  $\frac{1}{3\cdot 2^8(2\pi)^4}$  of the 
  coefficient $c_2$  in \rf{twoo}.
We conclude that  the 
coefficient of the 
$F^6$ term in the quantum 2-loop SYM effective action 
is the same as  in   the BI action \rf{www}.

\section{Summary and concluding  remarks}

In this paper we  have calculated the leading part of the 
planar two-loop contribution to the low-energy ${\cal N}=4$
$SU(N+1)$ SYM effective action  in the  abelian 
${\cal N}=2$   gauge  superfield background.
We used the 
formulation of ${\cal N}=4$ SYM theory in terms of ${\cal N}=2$
superfields in
harmonic superspace and the   background field
method.
 We have found that 
the relevant leading two-loop term   does not appear
from the 2-loop diagrams with 
 hypermultiplet and ghost propagators,
  so that 
the  result is given  entirely  by  the ${\cal N}=2$ gauge superfield 
contribution.\foot{While the $ F^6$ term   thus appears to be  
generated only by
the $\N=2$ SYM sector, the role of $\N=4$  
supersymmetry  is still important: 
it ensures the  cancellation of additional 
contributions to $F^6$ in which the harmonic derivative 
$D^{++}$ does not act on the ``box'' operator \rf{24a}.
Thus the result for the $F^6$ term in pure $\N=2$ SYM theory is 
expected to be
different from the $\N=4$ SYM one.}

The  calculation of the two-loop low-energy correction
we have described    is     a good 
illustration of the efficiency of the harmonic superspace approach
to computing the  effective action in ${\cal N}=4 $  and  
${\cal N}=2 $ supersymmetric gauge theories.  

The correction we  have calculated is the 
${\cal N}=2$ supersymmetrization of the $F^6/|X|^8$  term 
depending on a  constant $U(1)$ gauge field strength  and
 constant
scalar field 
background. This term does not appear in the 1-loop
approximation.  We have found that 
the large $N$  part of
its 2-loop coefficient  coincides 
exactly 
with that  of the  corresponding  term 
 in the expansion of the Born-Infeld action describing 
the  supergravity 
interaction between a stack of $N$ D3-branes and a  parallel 
 D3-brane  probe carrying constant $F_{mn}$  background.
Since,  in general,  the coefficient of this 
term in the  SYM effective action  could  be a non-trivial 
function of 't Hooft coupling (receiving corrections also from 
third  and higher loop orders), the agreement 
of the 2-loop coefficient 
with  the 
supergravity  expression 
(which, according to the AdS/CFT  correspondence, 
  should be describing the large-coupling 
  limit of the SYM theory)
is a strong indication of the  existence  of a  
non-renormalization 
theorem for this term (at least in the planar limit).

In section 1.2 we  proposed a  conjecture on how the correspondence
between the 
low-energy SYM  effective action and the   D3-brane  supergravity
interaction 
potential  can be extended to higher-order terms.
 Higher-dimensional $F^{2l+2}/|X|^{4l}$ terms in the SYM effective
 action
 should be  combinations of bosonic parts of several $\N=1$ (or
 $\N=2$)
 super-invariants. Only one of them 
 (for each $l$) 
 should have ``protected'' 
 coefficient which receives contribution only from  the $l$-th loop
 order. This term (its planar part)
 should thus 
 be the only one at given order that 
  survives  in the  strong coupling limit.
 It is this term that should  then  match onto  the 
 corresponding structure  in the 
 expansion of the Born-Infeld action, 
 in agreement with the expectation based on 
  AdS/CFT philosophy. 
 

 There are several possible  extensions
 of the present work   
  that may help to
 clarify the situation with the higher-order terms and may 
 provide some  evidence supporting  the  validity 
 of the above  conjecture:

 (i) Compute the  abelian $F^8$ term in the 
 2-loop SYM effective  action
 to demonstrate that it indeed   contains only the same
 ``unprotected"
 invariant $(F^8)_2$ as the 1-loop action \rf{yumi}
 -- the second  ``protected'' invariant  $(F^8)_1$
 should appear only at the {\it 3-loop} order.

 (ii) More generally, compute the full non-linear
 expression for the 2-loop SYM effective action in constant
 abelian
 $F_{mn}$ background, 
 i.e. the $D=4, \N=4$ SYM analog
 of the 2-loop Schwinger action in  quantum
 electrodynamics \ci{ritus}, or 
 the 1+3 dimensional counterpart
  of the  1+0  dimensional SYM result of \ci{bbb}.
  A comparison  of the corresponding 2-loop function of 
 the  $F_{mn}$ eigen-values  
 $\ff_1,\ff_2$ with
  the 1-loop  function  \rf{exx},\rf{yumi} should 
  be useful for identifying 
   $F^n$ invariants that  have ``unprotected" coefficients, 
   i.e. that appear in both 1-loop and 2-loop effective actions.

(iii)  Consider the SYM effective action 
in  a {\it non}-abelian $F_{mn}$ background and 
compute
   the 1-loop and 2-loop coefficients of the
 ``second'' non-abelian $F^6$ invariant
  (``${\rm tr}   ( F^4[F,F])$", see \ci{ch3}) 
   to confirm that its coefficient
   indeed  gets renormalized.

  (iv)  It  is well known that  
  the full non-abelian  1-loop $F^4$ term
   in the $\N=4$ SYM effective action -- 
   Str$[F^4 - {1\ov 4}(F^2)^2]$ can be obtained
  by taking the $\a'\to 0$ limit in the superstring
  partition function 
  on the annulus \ci{MT}; it would be of interest to give a
 similar string-theoretic 
  derivation  for the two-loop $F^6$ term
  (see in this connection   \ci{stri}).\foot{It would 
   also be useful
    to understand how to organize the  calculation
    of the 2-loop $F^6$ term directly in 
    10-dimensional  terms,
    making it possible to carry out the calculation in any 
    $1+n$ dimensional  reduction  of the $D=10, \N=1$ SYM theory.
    One would  then need to address the   issues of the 
    UV and IR cutoff dependence 
     which  were  absent in the $D=4$ case. }

  (v) Another useful generalization would be to
  repeat the calculation of the $F^6$ term in 
  other superconformal $\N=2$ theories,  corresponding, e.g.,
  to orbifold  versions  of the AdS/CFT    correspondence
 \ci{KS}. This would 
   allow one to check whether the  relation  between the 
 subleading $F^6$ 
 interactions on the supergravity and the SYM sides is still
 holding  in the less supersymmetric  situation
 (i.e.  whether  the conjectured 
  non-renormalization of the abelian 
 $F^6$ term is indeed  depending  on the 
 full $\N=4$ supersymmetry).

\bigskip \bigskip 
\noindent
{\bf Acknowledgements} \hfill\break
We would like to thank 
  S.M. Kuzenko for  many 
useful discussions and criticism 
 through  the course  of this work.
I.L.B. is  very grateful 
  to S.J. Gates for a discussion of some aspects of
supergraph evaluation and    for  the hospitality 
at the University of Maryland  during the final stage of this 
work.
 A.A.T. is grateful to W. Taylor for a
 useful discussion. 
 I.L.B. and A.A.T. acknowledge 
the support of the NATO collaborative research grant
 PST.CLG 974965.
The work by I.L.B. was  also supported in part by the RFBR grant
\symbol{242} 99-02-16617, and 
by the INTAS  grants 99-1590,  00-0254.
The work by A.Yu.P. was supported by FAPESP, project
No. 00/12671-7.
The work of A.A.T. was  partially supported by the DOE grant
  DE-FG02-91ER40690 and also by the PPARC SPG  00613,  INTAS 99-1590
 and  CRDF RPI-2108 grants. 

 
\renewcommand{\thesection}{}
\renewcommand{\theequation}{A.\arabic{equation}}
\setcounter{equation}{0}
\section{Appendix A}

Here we discuss the 
relation between the 
${\cal N}=2$ and $\cal N$=1 
supersymmetric 
forms of $F^6$ term.
Let us start with 
the ${\cal N}=2$ form of the $F^6$ correction
(see \rf{twoo}) 
\bea
\label{fo1}
S_6=\int d^{12}z\frac{1}{\bar{\W}^2}\log\frac{\W}{\m}D^4
\log\frac{\W}{\m}\  +  h.c. \ . 
\eea
Using that 
\bea
\int d^{12}z=\int d^4 x d^4\q_1\frac{1}{16} \bar{D}^2_2 D^2_2 
\eea
we get
\bea
S_6=\frac{1}{16}\int d^8z\bar{D}^2_2
D^2_2( \frac{1}{\bar{\W}^2}\log\frac{\W}{
\m}D^2_2 D^2_1\log\frac{\W}{\m}) + h.c. \ . 
\eea
Since $D^2_2 D^2_2=0$ this is equal to
\bea
S_6=\int d^8z
\bar{D}^2_2 D^2_2 (\frac{1}{\bar{\W}^2}\log\frac{\W}{\m}) |\ 
D^2_2 D^2_1\log\frac{\W}{\m}| + h.c. \ . 
\eea
Here the symbol $|$ denotes the value at $\q_2=0$.
The  chirality of $\W$ implies 
\bea
\bar{D}^2_2 D^2_2 ( \frac{1}{\bar{\W}^2}\log\frac{\W}{\m})=
\bar{D}^2_2 (\frac{1}{\bar{\W}^2}) D^2_2 (\log\frac{\W}{\m})\ . 
\eea
To express this in $\cal N$=1 form we use
 (see eq.(3.22) in \cite{bkt}; by  definition, 
$D^2_{\a}{\W}=2iW_{\a}$) 
\bea
\label{1}
D^2_2 D^2_1\log\frac{\W}{\m}|=4\frac{D^2 W^2}{\Phi^2}\ , 
\ \ \ \ \ \ \  
\bar{D}^2_2 \frac{1}{\bar{\W}^2}|=-
24\frac{\bar{W}^2}{\bar{\Phi}^4}\ , 
\ \ \ \ \ \  D^2_2\log\frac{\W}{\m}|=4\frac{W^2}{\Phi^2} \ . 
\eea
We thus finish with the following 
expression for the $\cal N$=1 form of $S_6$ 
\bea
S_6=-24\int d^8 z \frac{W^2\bar{W}^2 D^2 W^2}{|\Phi|^8} \ + h.c. \ . 
\eea

\renewcommand{\theequation}{B.\arabic{equation}}
\setcounter{equation}{0}
\section{Appendix B}
Here we shall  discuss the 
 calculation of the integral over the harmonics
in the expression for two-loop low-energy effective action 
(\ref{theres}).
Any integral of a function of  the 
  harmonics can be decomposed
into a sum of integrals of products, so 
  let us first 
describe the calculation of the 
 integral of an arbitrary product of
harmonics, e.g.,  
\bea \label{ih}  H =\int du \ u^+_{i_1}\ldots
u^+_{i_n} u^-_{j_1}\ldots u^-_{j_n}\ .  \eea
Here
$i_1,\ldots,i_n,j_1,\ldots,j_n=1,2$ are the 
$SU(2)$ harmonic indices taking values 1,2.
This expression contains
equal number of $u^+$ and $u^-$ harmonics,
since  otherwise  the integral is zero. 
 Any such  integral can be
calculated exactly using the 
 the general formalism  of \cite{GIKOS}, 
in particular, the properties 
\bea
\label{pr1}
& &\int du =1\ , \\
\label{pr2}
& &u^{+i}u^-_i=u^+_1u^-_2-u^-_1u^+_2=1\ , \ \ \ \ \ \ \ \ 
u^+_iu^-_j-u^+_ju^-_i=\epsilon_{ij}\ , \\
\label{pr3}
& &\int du u^{(+i_1}u^{+i_2}
\ldots u^{+i_n}u^{-j_1}\ldots u^{-j_m)}=0\ .  
\eea
The first relation defines the normalization of the integral, 
the  second expresses the fact that 
the determinant of a  matrix  of the harmonics
$||u^{\pm}_i||=\left(\begin{array}{cc}
u^+_1 & u^+_2\\
u^-_1 & u^-_2
\end{array}
\right)$
is equal to one, and the 
third means that the integral from the 
symmetrized product of
harmonics is equal to zero.

We  first  expand the product of harmonics 
 into a sum of terms symmetric
and antisymmetric in any pair of indices,  and 
then  use the  property (\ref{pr2})
which allows to replace the term $u^+_1u^-_2-u^+_2u^-_1$ by 1.
We finish with the 
 integral  which
contains the 
 sum of products of epsilon symbols with all possible ($n!$)
permutations
of indices $i_1,\ldots,i_n,j_1,\ldots,j_n$,
and the 
  sum of various
symmetrized products of harmonics. Due to (\ref{pr3}) the 
integral of any symmetrized product of harmonics gives zero. 
As a
result,
a non-vanishing 
 contribution comes from 
 the  sum of products of epsilon
symbols with different orders  of the indices. 
Since 
the expression
(\ref{ih}) is symmetric with respect to permutations 
of the 
indices $i_a$ as well as the indices $j_b$
($a,b=1, \ldots,  n$),
after (anti)symmetrization it
should  have the 
 factor $\frac{1}{(n!)^2}$.
 Then, the number of epsilon
symbols arising from 
 antisymmetrization is $n$ (the terms with less
number of epsilon symbols contain symmetric products of
harmonics,  and 
their integral  is zero (\ref{pr3})), and
there are $n!$ equivalent arrangements of these epsilon symbols.
Therefore, 
we  get the  combinatoric factor 
$\frac{1}{n!}$.  Also, the  (anti)symmetrization of any pair
of indices is accompanied  by   $1/2$, so that 
$n$ pairs of them require the factor
$\frac{1}{2^n}$.  As a result, the integral 
 (\ref{ih}) is equal to
\bea
\label{result} H=\frac{1}{2^n n!}\sum_{l=1}^n\prod_{m=1}^n
\epsilon_{i_m j_l}\ .  \eea
To compute the integral
in  (\ref{theres}) we note  that \ 
$({\cal D}^+(v){\cal W})^2({\cal D}^+(w){\cal W})^2 $ can  be
represented as\ \ 
$v^+_iv^+_jw^+_kw^+_l
{\cal D}^{i\a}{\cal W}{\cal D}^j_{\a}{\cal W}
{\cal D}^{k\b}{\cal W}{\cal D}^l_{\b}{\cal W}$. Then
the only non-zero
term in \\
  ${\cal D}^{i\a}{\cal W}{\cal D}^{j\a}{\cal W}
{\cal D}^{k\b}{\cal W}{\cal D}^{l\b}{\cal W}$ is
(due to anticommutativity 
of 
${\cal D}^{i\a}{\cal W}$-factors)\\ 
${\cal D}^{1\a}{\cal W}{\cal D}^1_{\a}{\cal W}
{\cal D}^{2\b}{\cal W}{\cal D}^2_{\b}{\cal W}$. A straightforward
calculation shows that $$
v^+_iv^+_jw^+_kw^+_l
{\cal D}^{i\a}{\cal W}{\cal D}^j_{\a}{\cal W}
{\cal D}^{k\b}{\cal W}{\cal D}^l_{\b}{\cal W}=
({\cal D}^1{\cal W})^2({\cal D}^2{\cal
W})^2(v^{+1}w^{+2}+v^{+2}w^{+1})^2 \ . $$
It is convenient to introduce  the notation
$$ v^{+1}w^{+2}+v^{+2}w^{+1}= 
d_{ij}v^{+i}w^{+j} \ , 
\ \ \ \ \ \ \ \ \ \ \ \  d_{ij}=\left(\begin{array}{cc}
0&1\\
1&0
\end{array}
\right)\ . 
$$
Using the relation \cite{GIKOS}:
\bea
(v^+w^+)(v^-w^-)=1+(v^+w^-)(v^-w^+)
\eea
we can  express (\ref{theres}) in the form
\bea
\label{theres2}
\Gamma^{(2)}&=&  \frac{1}{48(4\pi)^4}g^2_{\rm YM} N^2 \big[\int d^{12}z
\frac{1}{({\cW}\bar{\cW})^4} \bar{\cW}^2 
({\cal D}^1{\cal W})^2({\cal D}^2{\cal
W})^2  + h.c.\big] \nonumber\\&\times&
\int dv dw (d_{ij}v^{+i}w^{+j})^2
(v^-w^-)^2\Big(1-4(v^+w^-)(v^-w^+)+(v^+w^-)^2(v^-w^+)^2\Big)
\eea
Then the  products of the  harmonics in (\ref{theres})
 can be represented as:
\bea
(v^{+}w^{-})(v^{-}w^{+})&=&
\epsilon_{ij}\epsilon_{kl}v^{+i}w^{-j}v^{-k}w^{+l}\nonumber\\
{((v^+w^-)(v^-w^+))}^2
&=&\epsilon_{ij}\epsilon_{kl}\epsilon_{mn}\epsilon_{pq}
v^{+i}w^{-j}v^{+k}w^{-l}v^{-m}w^{+n}v^{-p}w^{+q}\\
{((v^+w^-)(v^-w^+))}^3&=&\epsilon_{ij}\epsilon_{kl}\epsilon_{mn}\epsilon_{pq}
\epsilon_{ab}\epsilon_{cd}
v^{+i}w^{-j}v^{+k}w^{-l}v^{+a}w^{-b}
v^{-m}w^{+n}v^{-p}w^{+q}v^{-c}w^{+d}
\ .  \nonumber\eea
Substituting   these expression into (\ref{theres2}) we get 
\bea
\Gamma^{(2)}&=& \frac{1}{48(4\pi)^4} g^2_{\rm YM} N^2 \big[ \int d^{12}z
\frac{1}{({\cW}\bar{\cW})^4}
\bar{\cW}^2({\cal D}^1{\cal W})^2({\cal D}^2{\cal
W})^2  + h.c. \big]    \nonumber\\&\times&
\int dv dw
(d_{ij}v^{+i}w^{+j})(d_{kl}v^{+k}w^{+l})(\e_{mn}v^{-m}w^{-n})
(\e_{pq}v^{-p}w^{-q})\nonumber\\&\times&
\Big[1-4\e_{ab}\e_{cd}v^{+a}v^{-c}w^{-b}w^{+d}\nonumber\\&+&
\e_{ab}\e_{cd}\e_{rs}\e_{tu}v^{+a}v^{+c}v^{-r}v^{-t}w^{-b}w^{-d}w^{+s}w^{+u}
\Big]
\eea
The integrand here is the 
sum of three terms.  Using  (\ref{result})
we find for  the first term 
\bea \label{b1} & &\int dv
dw (d_{ij}v^{+i}w^{+j})(d_{kl}v^{+k}w^{+l})(\e_{mn}v^{-m}w^{-n})
(\e_{pq}v^{-p}w^{-q})\nonumber\\&=&
d_{ij}d_{kl}{\e}_{mn}{\e}_{pq}
\left(\frac{1}{2^2 2!}\right)^2\nonumber\\&\times&
({\e}^{im}{\e}^{kp}+{\e}^{km}{\e}^{ip})
({\e}^{jn}{\e}^{lq}+{\e}^{jq}{\e}^{ln})=\frac{1}{16}\ , 
\eea
for second term 
\bea
\label{b2}
&-&4
\int dv dw
(d_{ij}v^{+i}w^{+j})(d_{kl}v^{+k}w^{+l})(\e_{mn}v^{-m}w^{-n})
(\e_{pq}v^{-p}w^{-q})
\e_{ab}\e_{cd}v^{+a}v^{-c}w^{-b}w^{+d}\nonumber\\&=&
-4d_{ij}d_{kl}{\e}_{mn}{\e}_{pq}{\e}_{ab}{\e}_{cd}
\left(\frac{1}{2^3 3!}\right)^2\nonumber\\&\times&
({\e}^{im}{\e}^{kp}{\e}^{ac}+{\e}^{im}{\e}^{kc}{\e}^{ap}+
{\e}^{km}{\e}^{ip}{\e}^{ac}+
{\e}^{km}{\e}^{ic}{\e}^{ap}
+{\e}^{am}{\e}^{kp}{\e}^{ic}+{\e}^{am}{\e}^{ip}{\e}^{kc})
\nonumber\\&\times&
({\e}^{jn}{\e}^{lq}{\e}^{db}+{\e}^{jn}{\e}^{lb}{\e}^{dq}+
{\e}^{ln}{\e}^{jq}{\e}^{db}+
{\e}^{ln}{\e}^{jb}{\e}^{dq}
+{\e}^{dn}{\e}^{jq}{\e}^{lb}+{\e}^{dn}{\e}^{jb}{\e}^{lq})=\frac{15}{32}
\eea
and for the third  term  
\bea
\label{b3}
& &\int dv dw
(d_{ij}v^{+i}w^{+j})(d_{kl}v^{+k}w^{+l})(\e_{mn}v^{-m}w^{-n})
(\e_{pq}v^{-p}w^{-q})
\e_{ab}\e_{cd}v^{+a}v^{+c}w^{-b}w^{-d}\nonumber\\&\times&
{\e}_{rs}{\e}_{tu}
v^{-r}v^{-t}w^{+s}w^{+u}
\nonumber\\&=&
d_{ij}d_{kl}{\e}_{mn}{\e}_{pq}{\e}_{ab}{\e}_{cd}{\e}_{rs}{\e}_{tu}
\left(\frac{1}{2^4 4!}\right)^2\nonumber\\&\times&
\left({\e}^{im}{\e}^{kp}{\e}^{ar}{\e}^{ct}+
{\e}^{im}{\e}^{kp}{\e}^{at}{\e}^{cr}
+{\e}^{im}{\e}^{kt}{\e}^{ar}{\e}^{cp}+...\right)
\nonumber\\
&\times&
\left({\e}^{jn}{\e}^{lq}{\e}^{sb}{\e}^{ud}+{\e}^{jn}{\e}^{lq}{\e}^{sd}{\e}^{ub}+
{\e}^{jn}{\e}^{ld}{\e}^{sb}{\e}^{uq}+...\right)
\nonumber\\
&=&
\frac{15}{32}
\eea
The sum of (\ref{b1}),(\ref{b2}),(\ref{b3}) is equal to 
1 and thus  we  finish with the expression  (\ref{res0}).


\end{document}